\documentclass[aps,prd,preprint,nofootinbib]{revtex4}
\usepackage{}
\usepackage{amsfonts}
\usepackage{epsfig}
\usepackage{graphicx}
\newcommand{\bqa}{\begin{eqnarray}}
\newcommand{\eqa}{\end{eqnarray}}

\begin{document}
\date{June 14, 2011}
\title{$W$ Boson Inclusive Decays to Quarkonium and $B_c^{(*)}$ Meson at the LHC\\[7mm]}

\author{Cong-Feng Qiao$^{a,b}\footnote{Email address: ~qiaocf@gucas.ac.cn}$,
 Li-Ping Sun$^{a}$\footnote{Email address: ~sunliping07@mails.gucas.ac.cn},
 De-Shan Yang$^{a}$\footnote{Email address: ~yangds@gucas.ac.cn},
 Rui-Lin Zhu$^{a}$\footnote{Email address: ~zhuruilin09@mails.gucas.ac.cn}} \affiliation{$a)$ College of
Physical Sciences, Graduate University of the Chinese Academy of
Sciences \\ YuQuan Road 19A, Beijing 100049, China}
\affiliation{$b)$ Theoretical Physics Center for Science Facilities
(TPCSF), CAS\\
 YuQuan Road 19B, Beijing 100049, China}

~\vspace{0.9cm}

\begin{abstract}
In this paper, the production rates of quarkonia $\eta_c$, $J/\psi$,
$\eta_b$, $\Upsilon$ and $B_c^{(*)}$ mesons through $W^+$ boson decay
at the LHC are calculated, at the leading order in both the QCD
coupling $\alpha_s$ and in $v$, the typical velocity of the heavy
quark inside of mesons. It shows that a sizable number of quarkonia
and $B_c^{(*)}$ mesons from $W^+$ boson decay will be produced at the LHC.
Comparison with
the predictions by using quark fragmentation mechanism is also
discussed. Results show that, for the charmonium production through
$W^+$ decay, the difference between predictions by the fragmentation
mechanism and complete leading order calculation is around $3\%$,
and it is insensitive to the uncertainties of theoretical
parameters; however, for the bottomonium and $B_c^{(*)}$
productions, the difference cannot be ignored as the fragmentation
mechanism is less applicable here due to the relatively large ratio
$m_b/m_w$.
\vspace {7mm}

\noindent {\bf PACS number(s):} 13.38.Be, 13.85.Ni, 14.40.Pq,
12.38.Bx.
\end{abstract}
\maketitle

\section{Introduction}

As the only charged gauge boson in the Standard Model (SM) whose mass
is generated through the electroweak symmetry spontaneous breaking
(EWSB), the precise measurements on $W$ boson's mass and decay width
offer a way to distinguish the different EWSB mechanisms, and
effectively put a stringent constraints on the Higgs boson mass and
New Physics beyond the SM\cite{Phy}. Furthermore, the study of the
productions of $W$ bosons can be a probe for searching New Physics
since the possible new particles may decay to $W$ bosons. By reason
of that, recently, there have appeared several papers discussing ``W boson
physics'', at Tevatron, ILC and LHC,  respectively \cite{Tev}.

Before the LHC era, only two rare decays:
$W^+\rightarrow\pi^+\gamma$ and $W^+\rightarrow D_s^+\gamma$, were
experimentally studied, and the respective upper limits for their
branching fractions are $8\times 10^{-5}$ and $1.3\times 10^{-3}$
\cite{Pdg}\footnote{Recently, the CDF collaboration 
just set a new $95\%$ confidence level upper limit on the relative branching fraction
${\Gamma(W^+\to \pi^{+}\gamma)}/{\Gamma(W^{+}\to e^{+}\nu)}$ at $6.4\times 10^{-5}$ which
is a factor of 10 improvement over the previous limit \cite{Aaltonen:2011nn}.}. The LHC has been running for two years. Due to its high
luminosity, the huge number of $W$ bosons are and will be produced.
According to the estimations in \cite{Phy,Coo,Jon}, the production
cross section of $W$ boson is around $\sim10^2$ nb at the LHC with
the center energy 7 TeV or 14 TeV. It is expected that hundreds of
million $W$ bosons will be produced per year at the LHC with the
conservative expectation of the luminosity $\sim
10^{32}\mathrm{cm}^{-2}\mathrm{s}^{-1}$ (which is twp orders of
magnitude smaller than the desired LHC luminosity $\sim
10^{34}\mathrm{cm}^{-2}\mathrm{s}^{-1}$). This makes the LHC also a
$W$-boson factory which facilitates the precise experimental study
on $W$ boson physics, especially $W$ boson rare decays.

  In this article, we will discuss a class of $W$ boson inclusive decays
to the S-wave quarkonium and $B_c$ meson, such as $\eta_c$, $J/\psi$, $\eta_b$,
$\Upsilon$, $B_c$ and $B_c^*$. Because $W$ boson is so heavy and
participates merely the electroweak interactions, the study on $W$
boson inclusive decays to quarkonia and $B_c^{(*)}$ mesons\footnote{As in\cite{Fra,Cha,Wu}, we treat $B_{c}^{(*)+}$ as 
a non-relativistic bound state of $c$ and $\bar b$ in this paper.}, offers another great place for
the precise test of the perturbative quantum chromodynamics (PQCD)
and the quarkonium production mechanism. As a quarkonium is presumed
to be a non-relativistic bound state of heavy quark and anti-quark,
a wonderful theoretical tool to deal with the processes involving
quarkonium is non-relativistic quantum chromodynamics (NRQCD),
in which the low-energy interactions are organized by the expansion
in $v$, the typical relative velocity of the heavy quark and
anti-quark inside of quarkonium\cite{Bod}.
 The inclusive production
cross sections can be written as the product of the perturbatively
calculable short-distance coefficients and the non-perturbative
NRQCD matrix elements. At the leading order in $v$, the only
non-perturbative NRQCD matrix element of the S-wave quarkonium is
proportional to the Schr\"odinger wave function at the origin squared.

On the other hand, since the $W$ boson mass $m_w$ is much greater
than the heavy quark masses $m_{b,c}$, the parton fragmentation may
dominate the $W$ boson inclusive decays to quarkonia and $B_c^{(*)}$ mesons. The pioneer
works in this field were done two decades ago, the universal
fragmentation functions for different quarkonium and $B_c$ meson were given in
\cite{Fra,Bra,Braa}. By these universal fragmentation functions, the
direct production rate of quarkonium and $B_c$ meson at large transverse momentum in
any high-energy process can be approximated to be the product of the
parton-level production rate and the universal fragmentation
probability, and the corresponding computation is much easier than
that done by within the NRQCD framework. In Ref.\cite{Fra,Bra,Bar},
for the $Z^0$ inclusive decays to the S-wave charmonium, the authors
did find great agreement between the results obtained by the parton
fragmentation approximation and the complete leading order PQCD
calculation.

However, in some processes, the fragmentation may not be the leading
process, but main process when a certain condition of the
fragmentation mechanism is not sufficiently fulfilled.  In
Ref.\cite{Qiao} it was found that the fragmentation contribution is
important but not dominant for top quark decays into quarkonium.
Thus, it is worth examining if the fragmentation mechanism works
in $W$ boson inclusive decays to quarkonium and $B_c$ meson.

In the following sections, we present the complete leading order
calculation of $W^+$ boson decays to quarkonium $\eta_c$, $J/\psi$,
$\eta_b$, $\Upsilon$ and $B_c^{(*)+}$ mesons; then we compare our results
with that obtained by the fragmentation approximation; finally we
discuss the theoretical uncertainties in our calculation.

\begin{figure}
 \centering
\includegraphics[width=0.50\textwidth]{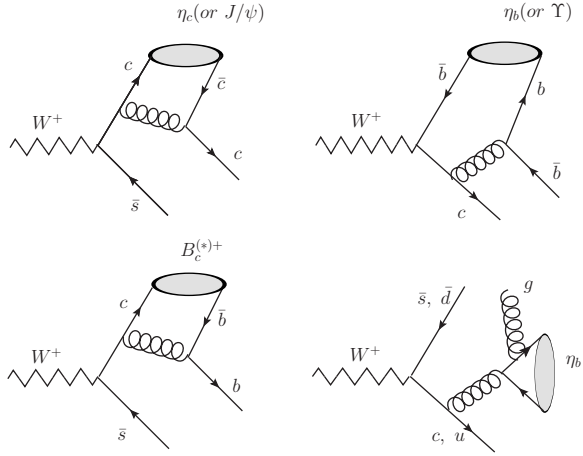}
\caption{The typical Feynman diagrams for quarkonia and $B_{c}^{(*)}$ productions from $W$ decays. Some diagrams with
the different gluon attachment are not shown.}\label{fig:etac}
\end{figure}

\section{Formalism}
Some typical Feynman diagrams for quarkonia and $B_c^{(*)}$ mesons hadronic production through $W^+$ decay 
are shown in Fig.  \ref{fig:etac}. We will calculate them at the leading order in $\alpha_{s}$ and $v$.
In NRQCD, a non-relativistic bound state of heavy quarks $Q$ and $\bar Q^{\prime}$
is considered as an expansion of a series Fock states. The leading Fock
state for the S-wave hadron is constructed as
\begin{eqnarray}
&&\vert H[^{2s+1}S_{J}](p,\lambda)\rangle\nonumber\\
&=&
{\sqrt{2 m_{H}}}\sum\limits_{i,j,\lambda_{1}\lambda_{2}}\frac{\delta_{ij}}{\sqrt{2N_{c}}}C(J,\lambda;\frac{1}{2},\lambda_{1},\frac{1}{2},\lambda_{2})
\nonumber\\
&&\times\int\frac{d^{3}\textbf{p}_{Q}}{\sqrt{2 E_{Q}2E_{Q^{\prime}}}}\tilde{\Psi}_{^{2s+1}S_{J}}(\textbf{p}_{Q})
\vert Q_{i,\lambda_{1}}(p_{Q})\bar Q^{\prime}_{j,\lambda_{2}}(p_{Q^{\prime}})\rangle\,,
\end{eqnarray}
where $m_{H}$ denotes the mass of hadron and all the states are relativistically normalized.
Here $p$ is
momentum of hadron, $i,j$ stand for the color indices which run from $i,j=1,...,N_{c}$ with
$N_c=3$ for QCD, and $C(J,\lambda;1/2,\lambda_{1},1/2,\lambda_{2})$ is the C-G coefficient with $\lambda,\lambda_{1}$ and $\lambda_{2}$ being
the third components of spin indices. The non-perturbative parameters $\tilde{\Psi}_{^{2s+1}S_{J}}(\textbf{p}_{Q})$ is
 the Schr\"{o}dinger wave functions in momentum space. Thus, to project out the amplitude for the S-wave hadronic states 
from the complete patron level amplitude at the leading order of $v$, practically we do the following
replacements for the heavy quark $Q$ and -anti-quark $\bar Q^{\prime}$ spinors
\begin{eqnarray}\label{eq:projector}
v_{i}(p_{Q^{\prime}})\bar u_{j} (p_{Q})&\to&\frac{\Psi_{^{1}S_{0}}(0)}{2\sqrt{m_{Q}+m_{Q^{\prime}}}} \gamma^{5}({{p}\!\!\!\slash}+m_{Q}+m_{Q^{\prime}})
\frac{\delta_{ij}}{\sqrt{N_c}}
,\\
v(p_{Q^{\prime}})\bar u (p_{Q})&\to&\frac{\Psi_{^{3}S_{1}}(0)}{2\sqrt{m_{Q}+m_{Q^{\prime}}}}{\epsilon}\!\!\!\slash^*({{p}\!\!\!\slash}
+m_{Q}+m_{Q^{\prime}})\frac{\delta_{ij}}{\sqrt{N_c}}.
\end{eqnarray}
 Here $\epsilon^{\mu}$ is the polarization
vector of $^{3}S_{1}$ state, $\Psi_{^{2s+1}S_{J}}(0)$ is the Schr\"{o}dinger wave function at the
origin. At the heavy quark limit, 
we set $\Psi_{^{3}S_{1}}(0)=\Psi_{^{1}S_{0}}(0)$.
Throughout the paper, we adopt the relation $m_H=m_{Q}+m_{Q^{\prime}}$ since we 
are doing the LO calculation in $v$. Obviously, we have $m_{Q}=m_{Q^{\prime}}$ for quarkonia, 
and $m_{Q}=m_{b}$, $m_{Q^{\prime}}=m_{c}$ for
$B_{c}^{(*)}$. 

\subsection{$\quad W^{+}\rightarrow \eta_{c}({\rm or}~J/\psi)+c+\bar{s}$}
There are two Feynman diagrams for $W^{+}$ boson decay into a S-wave
charmonium state associated with $c$ and $\bar{s}$. 
Implementing the Feynman rules of the Standard Model and projectors in Eq.(\ref{eq:projector}), the amplitude
for $W^+\to\eta_c+c+\bar{s}$ is
\begin{equation}
M =
\frac{16\pi}{3}\frac{g\alpha_s V_{cs}}{2\sqrt{2}}\frac{\Psi_{\eta_c}(0)}{2\sqrt{6m_c}}
(A_1+A_2),
\end{equation}
where
\begin{eqnarray}
A_1&=&\bar{u}(p_6)\gamma^{\alpha}\gamma^5(2m_c+p\!\!\!\slash)\epsilon\!\!\!\slash_W
(1-\gamma^5)\nonumber\\
&&\frac{(-p\!\!\!\slash_3-p\!\!\!\slash_5-p\!\!\!\slash_6)}
{(p_3+p_5+p_6)^2}\gamma_\alpha v(p_5)\frac{1}{(p_3+p_6)^2}\,,
\\
A_2&=&\bar{u}(p_6)\gamma^{\alpha}\gamma^5(2m_c+p\!\!\!\slash)\gamma_\alpha\frac{(
m_c+p\!\!\!\slash_3+p\!\!\!\slash_4+p\!\!\!\slash_6)}{(p_3+p_4+p_6)^2-m_c^2}
\epsilon\!\!\!\slash_W\nonumber\\
&&(1-\gamma^5)v(p_5)\frac{1}{(p_3+p_6)^2}\,,
\end{eqnarray}
with $p$, $p_5$ and $p_6$ being the momenta of $\eta_c$ meson,
$\bar{s}$ quark and $c$ quark, respectively; $p_3$ and $p_4$ the
momenta of $\bar{c}$ quark and $c$ quark in $\eta_c$ meson;
 $V_{cs}$ the CKM matrix element;
 $g= e/\sin \theta_{{W}}$ with $\theta_{W}$ the Weinberg angle and $e$ the 
 unit electro-charge; $\alpha\equiv e^{2}/(4\pi)$
 and $\alpha_s=g_{s}^{2}/(4\pi)$ are the electromagnetic and the strong coupling constants,  respectively;
 $\epsilon_W$ the polarization vector of $W^+$ boson.
 In this paper, we set $u$, $d$ and $s$ quarks massless, and ignore
 the dependence on the relative momentum between quark and anti-quark in meson, and set $p_3=p_4=p/2$.
The spin-averaged partial decay width of $W^{+}\rightarrow
\eta_{c}+c+\bar{s}$ reads
\begin{eqnarray}
d\Gamma&=&\frac{1}{2^8m_w^3\pi^3}\sum\limits_{\mbox{spins}}
|M|^2ds_1ds_2,
\end{eqnarray}
where $m_w$ is the mass of $W$ boson, $s_1=(p+p_5)^2$ and
$s_2=(p_5+p_6)^2$. The explicit analytic expressions for the square
of the amplitude for $W^+\to \eta_c+c+\bar s$ is given in Appendix
A.

 For the complete leading order calculation of  $W^{+}\rightarrow
J/\psi+c+\bar{s}$, we just repeat the calculation that had been done
in Ref.\cite{Bar}. We find that there is a misprint in the term
$-8xy^3r^{\prime}M_W^2(1+r-r^{\prime}-r)^{-2}$ in $\sum |A_2|^2$ given in
Appendix B of \cite{Bar}, which should be corrected to
$-8xy^2r^{\prime}M_W^2(1+r-r^{\prime}-r)^{-2}$.

\subsection{$W^{+}\rightarrow \eta_{b}({\rm or}~\Upsilon)+X$}
$W^{+}\rightarrow \eta_{b}({\rm or}~\Upsilon)+c+\bar{b}$ is the leading order process of bottomonium production through $W^+$ decay in the expansion of
$\alpha_{s}$. However, such processes are the CKM-suppressed with the corresponding CKM factor $V_{cb}=A\lambda^{2}$ in the Wolfenstein parameterization where $A=0.813$ and $\lambda=0.225$. Numerically, the Wolfenstein parameter $\lambda\sim \alpha_{s}(2 m_{c})=0.26$.  Hence, we will consider some processes that are higher order in expansion of $\alpha_{s}$ but lower order in expansion of the Wolfenstein parameter $\lambda$. For instance, the process $W^{+}\rightarrow \eta_{b}+c+\bar s +g$ depicted by the last diagram  in Fig.  \ref{fig:etac} is associated with the factor $g_{s}^{3}V_{ud}\sim g_{s}^{3}\lambda^{0}$ which is numerically comparable to the factor $g_{s}^{2}V_{cb}\sim g_{s}^{2}\lambda^{2}$ accompanied with the process $W^{+}\rightarrow \eta_{b}+c+\bar{b}$. Similarly, we also consider $W^{+}\to  \Upsilon+c+\bar s+\gamma$ for $W$ decays to $\Upsilon$. Its amplitude is proportional to $V_{ud} e^{2}$, which is numerically comparable to the factor $g_{s}^{2}V_{cb}$ accompanied with $W^{+}\rightarrow \Upsilon+c+\bar{b}$.

 The calculation of $\eta_b$ production is similar to the
$\eta_c$ production described in the previous subsection. Here we
just show the results for the production of $\Upsilon$. The complete
amplitude for $W^+$ decay to $\Upsilon$ with $c$ and $\bar{b}$ is
\begin{equation}
M=\frac{16\pi}{3}\frac{g\alpha_s V_{cb}}{2\sqrt{2}}\frac{\Psi_{\Upsilon}(0)}{2\sqrt{6m_b}}(A_1+A_2),
\end{equation}
with
\begin{eqnarray}
A_1&=&\bar{u}(p_5)\gamma^{\alpha}\frac{(m_c+p\!\!\!\slash_3+p\!\!\!\slash_5+p\!\!\!\slash_6)}{(p_3+p_5+p_6)^2
-m_c^2}\epsilon\!\!\!\slash_W(1-\gamma^5)\epsilon\!\!\!\slash_\Upsilon^*\nonumber
\\
&&~~(2m_b+p\!\!\!\slash)
\gamma_\alpha v(p_6)\frac{1}{(p_3+p_6)^2},
\\
A_2&=&\bar{u}(p_5)\epsilon\!\!\!\slash_W(1-\gamma^5)\frac{(m_b-p\!\!\!\slash_3-p\!\!\!\slash_4-p\!\!\!\slash_6)}
{(p_3+p_4+p_6)^2-m_b^2}\gamma^{\alpha}\epsilon\!\!\!\slash_\Upsilon^*\nonumber
\\
&&~~(2m_b+p\!\!\!\slash)\gamma_
\alpha v(p_6)\frac{1}{(p_3+p_6)^2},
\end{eqnarray}
where $p$, $p_5$ and $p_6$ are the momenta of $\Upsilon$, $c$ quark
and $\bar{b}$ quark, respectively;  $p_3$ and $p_4$ are the momenta
of $b$ quark and  $\bar{b}$ quark in $\Upsilon $ meson;
$\epsilon_\Upsilon$ is the polarization vector of $\Upsilon$ boson.
The explicit expression for the square of amplitude is given in
Appendix B.
\begin{figure}
 \centering
\includegraphics[width=0.50\textwidth]{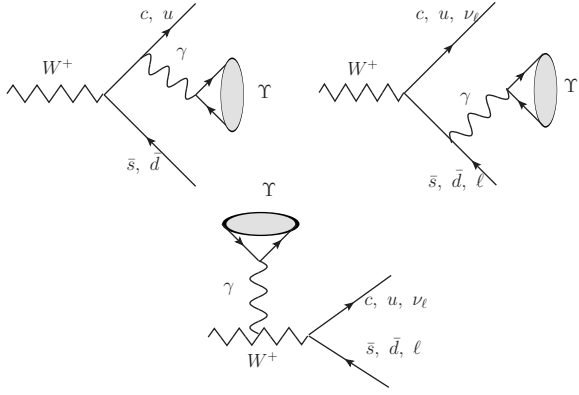}
\caption{Relevant electroweak Feynman diagrams which contribute in
the production of $\Upsilon$. } \label{fig:ewcon}
\end{figure}

Because of the CKM suppression, the electroweak contributions
depicted in Fig.  \ref{fig:ewcon} is required for the production of
$\Upsilon$ as we argued above\footnote{However, for that of $J/\psi$, the electroweak contribution
can be ignored in LO,
which is two orders of magnitude less than its QCD contribution.}.
Taking $W^{+}\rightarrow \Upsilon+u+\bar{d}$ for
example, the amplitude for this process is written as
\begin{equation}
M=\sqrt{3}ge^2V_{ud}\frac{\Psi_{\Upsilon}(0)}{2\sqrt{6m_b}}(A_1+A_2+A_3),
\end{equation}
with
\begin{eqnarray}
A_1&=&\frac{\mathrm{Tr}[\gamma_{\alpha}\epsilon\!\!\!\slash_\Upsilon^*(2m_b+p\!\!\!\slash)]}{9\sqrt{2}(p_3+p_4)^2}
\nonumber\\
&&\times\bar{u}(p_6)\gamma^{\alpha}\frac{p\!\!\!\slash_3+p\!\!\!\slash_4+p\!\!\!\slash_6}{(p_3+p_4+p_6)^2
}\epsilon\!\!\!\slash_W(1-\gamma^5)v(p_5)\,,
\\
A_2&=&-\frac{\mathrm{Tr}[\gamma_{\alpha}\epsilon\!\!\!\slash_\Upsilon^*(2m_b+p\!\!\!\slash)]}{18\sqrt{2}(p_3+p_4)^2}
\nonumber\\
&&\times\bar{u}(p_6)\epsilon\!\!\!\slash_W(1-\gamma^5)\frac{(-p\!\!\!\slash_3-p\!\!\!\slash_4-p\!\!\!\slash_5)}{(p_3+p_4+p_5)^2
}\gamma^{\alpha}v(p_5)\,,
\\
A_3&=&-\frac{\mathrm{Tr}[\gamma^{\alpha}\epsilon\!\!\!\slash_\Upsilon^*(2m_b+p\!\!\!\slash)]
\bar{u}(p_6)\gamma^{\nu}(1-\gamma^5)\epsilon^\mu_W v(p_5)}{6\sqrt{2}(p_3+p_4)^2((p_5+p_6)^2
-m_w^2)}\nonumber
\\
&&\times\Big((p_5+p_6-p)_{\mu}g_{\alpha\nu}-(2p_5+2p_6+p)_{\alpha}g_{\mu\nu}\nonumber
\\
&&~~+(p_5+p_6+2p)
_{\nu}g_{\alpha\mu}\Big)\,,
\end{eqnarray}
where $p$, $p_5$ and $p_6$ are the momenta of $\Upsilon$, $\bar{d}$
quark and $u$ quark, respectively; $p_3$ and $p_4$ are the momenta
of $b$ quark and  $\bar{b}$ quark in $\Upsilon $ meson. Here again
we set $p_3=p_4=p/2$ in the calculation.

$W^{+}\to \eta_{b}+ c+\bar s +g$ dominates the decay $W^{+}\to \eta_{b}+ X$. Its four-body decay amplitude is quite complicated, so
we do not present the complete analytic expressions here but the numerical results in the next section. 
The potential infrared-divergences arising from the regions where the momentum of
the gluon becomes soft vanish due to the ``color-transparency'', and the potential collinear-divergences arising from the region where the momentum of the gluon becomes collinear to the $\eta_{b}$ and $c$-quark (or $\bar s$ quark) are regulated by the $b$ quark mass. Thus, the decay width 
of $W^{+}\to \eta_{b}+ c+\bar s+ g$ is infrared-safe. 

\subsection{$\quad W^{+}\rightarrow B_c^{(*)+}+b+\bar{s}$ and  $W^{+}\rightarrow B_c^{(*)+}
+c+\bar{c}$}

There are two processes, namely $W^{+}\rightarrow
B_c^{(*)+}+b+\bar{s}$ and $W^{+}\rightarrow B_c^{(*)+} +c+\bar{c}$,
for $\bar{b}c$ bound states production in $W^+$ decay at the leading
order. $W^{+}\rightarrow B_c^{(*)+}+g+g$ and $W^{+}\rightarrow B_c^{(*)+}+\gamma$ can also 
contribute, however their contributions vanish at the large $m_{w}$ limit, and thus  
are almost two and three orders of magnitude lower than $W^{+}\rightarrow
B_c^{(*)+}+b+\bar{s}$ and $W^{+}\rightarrow B_c^{(*)+} +c+\bar{c}$ numerically. 

For $W^{+}\rightarrow B_c^{+}+b+\bar{s}$, there are two
Feynman diagrams and the corresponding amplitude is
\begin{equation}
M=\frac{16\pi}{3}\frac{g\alpha_s V_{cs}}{2\sqrt{2}}\frac{\Psi_{B_c}(0)}{2\sqrt{3(m_c+m_b)}}(A_1+A_2),
\end{equation}
with
\begin{eqnarray}
A_1&=&\frac{1}{(p_3+p_6)^2}\bar{u}(p_6)\gamma^{\alpha}\gamma^5(m_c+m_b+p\!\!\!\slash)\epsilon\!\!\!\slash_W(1-\gamma^5)\nonumber
\\
&&
\frac{(-p\!\!\!\slash_3-p\!\!\!\slash_5-p\!\!\!\slash_6)}{(p_3+p_5+p_6)^2}
\gamma_\alpha v(p_5),
\\
A_2&=&\frac{1}{(p_3+p_6)^2}\bar{u}(p_6)\gamma^{\alpha}\gamma^5(m_c+m_b+p\!\!\!\slash)\gamma_\alpha\nonumber
\\
&&~~\frac{(m_c+p\!\!
\!\slash_3+p\!\!\!\slash_4+p\!\!\!\slash_6)}{(p_3+p_4+p_6)^2-m_c^2}\epsilon\!\!\!
\slash_W(1-\gamma^5)v(p_5).
\end{eqnarray}
Here $p$, $p_3$, $p_4$, $p_5$ and $p_6$ are, respectively, the
momenta of $B_c^{+}$, $\bar{b}$ quark and $c$ quark in $B_c^{+}$ meson,
$\bar{s}$ quark and $b$ quark.
 The corresponding amplitude at the leading order
for $W^{+}\rightarrow B_c^{+}+c+\bar{c}$ is
\begin{equation}
M=\frac{16\pi}{3}\frac{g\alpha_s V_{cb}}{2\sqrt{2}}\frac{\Psi_{B_c}(0)}{2\sqrt{3(m_c+m_b)}}
(A_1+A_2),
\end{equation}
with
\begin{eqnarray}
A_1&=&\bar{u}(p_6)\gamma^{\alpha}\frac{(m_c+p\!\!\!\slash_4+p\!\!\!\slash_5+p\!\!\!\slash_6)}
{(p_4+p_5+p_6)^2-m_c^2}\epsilon\!\!\!\slash_W(1-\gamma^5)\nonumber\\
&&~~\gamma^5(m_c+m_b+p\!\!
\!\slash)\gamma_\alpha v(p_5)\frac{1}{(p_4+p_5)^2},
\\
A_2&=&\bar{u}(p_6)\epsilon\!\!\!\slash_W(1-\gamma^5)\frac{(m_b-p\!\!\!\slash_3-p\!\!\!\slash_4-p
\!\!\!\slash_5)}{(p_3+p_4+p_5)^2-m_b^2}\gamma^{\alpha}\nonumber\\
&&~~\gamma^5(m_c+m_b+p\!\!\!
\slash)\gamma_\alpha v(p_5)\frac{1}{(p_4+p_5)^2}\,,
\end{eqnarray}
where $p$, $p_3$, $ p_4$, $p_5$ and $p_6$ are, respectively, the
momenta of $B_c$, $\bar{b}$ quark and $c$ quark in $B_c$ meson,
$\bar{c}$ quark and $c$ quark.

Similarly, we can obtain the amplitudes for $B_c^*$ production.
In Appendix C, we detail the square of the amplitudes for both
$W^{+}\rightarrow B_c^{(*)+}+b+\bar{s}$ and $W^{+}\rightarrow
B_c^{(*)+} +c+\bar{c}$.

\section{Numerical Results and Analysis}
\subsection{Input parameters}

Now we can employ the above formalism to evaluate the decay widths
of $W^+$ boson to quarkonia and $B_c^{(*)+}$ mesons. We adopt the mass parameters for $c$ quark, $b$ quark and $W$ boson as follows
\[
 m_c=1.50\,\mathrm{GeV}\,,~~m_b=4.90\,\mathrm{GeV}~~{\rm and}~~m_w=80.4\,\mathrm{GeV}.
 \]
  The Schr\"{o}dinger wave functions
at the origin for  $J/\psi$ and  $\Upsilon$ are determined through their leptonic decay widths
$\Gamma_{ee}$, at the leading order in both $\alpha_s$ and $v$, which is \cite{Bod},
\[
|\Psi_{\psi (\Upsilon)}(0)|^2=\frac{m^2_{\psi (\Upsilon)} \Gamma_{ee}}{16\pi\alpha^2e_{c(b)}^2}.
\] 
Taking $\Gamma_{ee}^\psi=5.55~\mathrm{keV}$, $\Gamma_{ee}^\Upsilon=1.34~\mathrm{keV}$, $\alpha=1/137$ and $e_c=2/3$, we obtain
\[
|\Psi_{\psi}(0)|^2=0.0447~\mathrm{GeV}^3~~{\rm and}~~|\Psi_{\Upsilon}(0)|^2=0.403~\mathrm{GeV}^3.
\]

In principle, we can extract $\Psi_{\eta_{c}}(0)$ from the decay width of $\eta_{c}\to\gamma\gamma$. However, since
 the experimental data uncertainty is still large, which is $B(\eta_c\rightarrow \gamma \gamma)=(6.3\pm2.9)\times 10^{-5}$ \cite{Pdg}, 
the corresponding extraction of $\Psi_{\eta_{c}}(0)$ is not applicable in our numerical analysis. Instead, we set $\Psi_{\eta_{c}}(0)=\Psi_{J/\psi}(0)$, which is the consequence of the heavy quark spin symmetry in NRQCD at leading order $v$ \cite{Bod}. Similarly, we adopt $\Psi_{\eta_{b}}(0)=\Psi_{\Upsilon}(0)$. 

For $B_c^{(*)}$, we apply $|\Psi_{B_c}(0)|^2=|\Psi_{B_c^*}(0)|^2=0.1307(\mathrm{GeV})^3$ given in \cite{Wu} by using the Buchm\"{u}ller-Tye potential \cite{Eichten}.
 In addition, for consistency the leading order $\alpha_s$ running is adopted, i.e. 
 \[
 \alpha_s (\mu)=\frac{4\pi}{(11-2n_f/3)\ln (\mu^2/\Lambda^2_{QCD})},
 \]
  where we take $\Lambda^2_{QCD}=200\mathrm{MeV}$ and $n_{f}=4$ as in \cite{Wu}.
 We choose the typical renormalization scale $\mu=2m_c$ for charmonium and $B_c^{(*)}$ production, and $\mu=2m_b$ for bottomonium production, correspondingly $\alpha_s(2 m_c)=0.26$ and $\alpha_s(2 m_b)=0.19$.
 
  \begin{table*}
  \caption{Decay widths and branching fractions of
  quarkonium and $B_c$ meson through $W^+$ inclusive decays. Here the $q_i\bar{q_j}$ represents one kind of $c\bar{s}$, $b\bar{s}$,
  $c\bar{c}$, $c\bar{b}$, $u\bar{d}$, $\mu^+ \nu_\mu$ and $c\bar{s}g$; the $Q_i\bar{Q_j}$ one kind of $c\bar{s}$ and $c\bar{b}$. }\label{tab:width}
\begin{tabular*}{\textwidth}{@{\extracolsep{\fill}}ccc@{}}
\hline
 $W^+\rightarrow H q_i\bar{q_j}$ & $\Gamma(W^+\rightarrow H q_i\bar{q_j})
  $(keV) & $\Gamma(W^+\rightarrow H q_i\bar{q_j})/\Gamma(W^+\rightarrow Q_i\bar{Q_j})$\\
   \hline $W^+\rightarrow J/\psi c\bar{s}$ & $90.4$ & $1.27\times10^{-4}$ \\
     $W^+\rightarrow\eta_c c\bar{s}$ & $87.5$ & $1.23\times10^{-4}$ \\
    $W^+\rightarrow B_c^+ b\bar{s}$ & $6.3$ & $8.84\times10^{-6}$ \\
    $W^+\rightarrow B_c^{*+} b\bar{s}$ & $5.4$ & $7.61\times10^{-6}$ \\
   $W^+\rightarrow B_c^+ c\bar{c}$ & $0.4$ & $3.64\times10^{-4}$ \\
    $W^+\rightarrow B_c^{*+} c\bar{c}$& $0.5$ & $5.23\times10^{-4}$ \\
    $W^+\rightarrow\eta_b c\bar{b}$ & $0.015$& $1.41\times10^{-5}$ \\
    $W^+\rightarrow\Upsilon c\bar{b}$& $0.016$ & $1.49\times10^{-5}$ \\
    $W^+\rightarrow\Upsilon u\bar{d}$& $0.014$ & $$ \\
    $W^+\rightarrow\Upsilon \mu^+ \nu_\mu$& $0.013$ & $$ \\
    $W^+\rightarrow\eta_b c\bar{s}g$& 0.517$$ & $$ \\
  \hline
\end{tabular*}
\end{table*}

\subsection{Decay widths and feasibility at LHC}
By the input parameters given above, we list the partial widths of $W$ decays to the S-wave quarkonia and $B_c^{(*)}$ mesons in
 Table \ref{tab:width}. Employing these partial widths, one can estimate the event numbers of quarkonium production through $W$ decays at the LHC. Considering
that the LHC runs at the center-of mass energy
$\sqrt{s}=14\mathrm{TeV}$ with the luminosity $\mathcal
{L}_{p-p}=10^{34} \mathrm{cm}^{-2}\mathrm{s}^{-1}$, and the cross
section  $\sigma_{W^+}=10^2
 \mathrm{nb}$ at the LHC \cite{Jon}, the number of $W^+$ events per year is expected to be $3.07
 \times10^{10}$. Then we present the event rates of $W$ decays to quarkonia or $B_{c}^{(*)}$ in Table \ref{tab:ann}.

The most readily identifiable quarkonia are $J/\psi$ and $\Upsilon$, because their leptonic decays have clear signals and  
relatively large branching fractions. With $B(\psi\rightarrow \mu\mu)=5.93\%$,
 $B(\Upsilon\rightarrow \mu\mu)=2.48\%$ and $B(\Upsilon\rightarrow \tau\tau)=2.60\%$, we list the expected di-leptonic signals
from $J/\psi$ and $\Upsilon$ decays in Table \ref{tab:lep}. For experimental detection for such decays at the LHC, one has to look for the relevant events 
$pp\to W^{+} X$ with $W\to J/\psi(\mu^{+}\mu^{-})+2~{\rm jets}$. According to \cite{Gao}, the efficiency of reconstruction of $J/\psi$ from its dimuon decay channel is around $40\%$ for $pp$ collision at LHC experiments. However, to tag whether such $J/\psi$ event is really from $W^{+}$ decay, one must reconstruct $W^{+}$ from $\mu^{+}\mu^{-}+2~{\rm jets}$ event. Therefore, the additional information about relevant QCD background of such events could be crucial as well. 


The most promising decay channels for reconstruction $\eta_{c}$ are $\eta_{c}\to K\bar K \pi$ , $\eta \pi\pi$ and etc, which branching ratios are around a few percent. Thus, it should be possible to observe $W^{+}$ decays to $\eta_{c}$ at the LHC. Certainly, such measurements also suffers the difficulty of reconstruction of $W^{+}$ as mentioned above.

 $B_c^+\to J/\psi\pi^{+}$ is the most promising channels to identify $B_{c}^{+}$ at colliders. However, considering their small branching fractions and the efficiency of the event reconstruction, the measurements on $W$ decays to $B_{c}$ at the LHC would be quite difficult. With the similar reasoning, it would be barely possible to measure $W$ decays to $\eta_{b}$ at the LHC due to the lower branching fraction and the experimental difficulty for identifying $\eta_{b}$.

  \begin{table*}
  \caption{The expected annual event numbers of the quarkonium and $B_c$ meson from $W^+$ decay at the
  LHC with
$\sqrt{s}=14\mathrm{TeV}$ and $\mathcal {L}_{p-p}=10^{34}
\mathrm{cm}^{-2}\mathrm{s}^{-1}$.}\label{tab:ann}
\begin{tabular*}{\textwidth}{@{\extracolsep{\fill}}ccccccc@{}}
  \hline
  & $J/\psi$ & $\eta_c$ & ~~~$B_c^+$~~~ & ~~~$B_c^{*+}$ ~~~& ~~~$\eta_b$~~~ & ~~~$
  \Upsilon$~~~ \\ \hline
  Events number($\times10^4$) & ~~~133 ~~~& ~~~129~~~ & ~~~9.8~~~ & ~~~8.8~~~
  & ~~~1.5
  ~~~ & ~~~0.12~~~ \\
  \hline
\end{tabular*}
\end{table*}

\begin{table*}
  \caption{The corresponding annual leptonic events of $J/\psi$ and $\Upsilon$. }\label{tab:lep}\vskip 0.5cm
\begin{tabular*}{\textwidth}{@{\extracolsep{\fill}}cccccc@{}}
  \hline
  &\multicolumn{2}{c}{  $J/\psi$ }&  \multicolumn{3}{c}{$\Upsilon$}
     \\ \hline
  Decay channel& $e^+e^-$ & $\mu^+\mu^-$ & ~~~$e^+e^-$~~~ & ~~~$\mu^+\mu^-$ ~~~& ~~~$\tau^+\tau^-$~~~
   \\ \hline
  Events & ~~~7.9$\times10^4$ ~~~& ~~~7.9$\times10^4$ ~~~ & ~~~30~~~ & ~~~30~~~
  & ~~~
  31~~~ \\
  \hline
\end{tabular*}
\end{table*}

\subsection{Comparisons with the fragmentation mechanism}
Since $m_w\gg m_b\,,m_c$, one can argue that the fragmentation
mechanism may be the dominant contribution to the inclusive decay
rate of the $W^+$ into the quarkonium and $B_c$ meson which survives in the limit
$m_w/m_{b,c}\to \infty$. In the fragmentation mechanism, the hadron
$H$ with energy $E$ is produced by the fragment of a type $i$ parton
with energy $E/z$ ($z$ is the longitudinal momentum fraction of $H$
relative to type $i$ parton) directly from the decay of $W^+$ boson.
The possibility of $i\to H$ is presumed to be described by the
universal fragmentation function $D_{i\to H}(z)$. Putting all the
possible parton fragmentation together, the differential decay width
of $H$ inclusive production can be written as
\begin{eqnarray}\label{eq:frag}
&&d\Gamma(W^+\to
H(E)+X)\nonumber\\
&=&\sum\limits_i\int_0^1\,dz\,d\hat{\Gamma}(W^+\to
i(E/z)+X,\mu)D_{i\to H}(z,\mu)\nonumber\\
&&~~+{\cal O}\left(\frac{m_H}{m_w}\right),
\end{eqnarray}
where $d\hat{\Gamma}$ is the parton level differential decay width, and
the sum is over the parton type $i$ and the hadron $H$'s
longitudinal momentum fraction $z$. At the leading order of $\alpha_s$, $d\hat{\Gamma}$ does not
depend on the longitudinal momentum fraction $z$. Thus, the total
decay width turns to be
\begin{eqnarray}
&&\Gamma(W^+\to H(E)+X)=\sum\limits_i\,\hat{\Gamma}(W^+\to
i(E/z)+X)P_{i\to H}\,,\nonumber\\\\
&&P_{i\to H}\equiv\int_0^1\,dzD_{i\to H}(z,\mu)\,,
\end{eqnarray}
where $P_{i\to H}$ is the so-called fragmentation possibility.

After paying the price of ${\cal O}(m_H/m_w)$ power corrections,
Eq.(\ref{eq:frag}) has a number of advantages in calculation. The
parton level differential decay width $d\hat{\Gamma}$ is easy to be
calculated and dependent only on the typical short-distance scale
$m_w$. The fragmentation functions are universal and dependent on
the hadronization scale, and therefore, they can be either extracted
from the experimental measurements or calculated from certain
phenomenological models. Fortunately, the fragmentation functions
for the S-wave quarkonium and $B_c$ meson can be calculated
perturbatively \cite{Bra,Braa}, together with all the
non-perturbative hadronization effects being parameterized into the
Schr\"odinger wave functions of the mesons at origin at the
non-relativistic limit.

 In calculation we take the following fragmentation probabilities from \cite{Bra,Braa}:
\begin{equation}
~~~~~P(c\rightarrow \psi) \;=\;
 {32 \alpha_s^2(2 m_c) |\Psi_\psi(0)|^2 \over 27  m_c^3}
\left( {1189 \over 30} - 57 \ln 2 \right),
\end{equation}
\begin{equation}
~~~~~P(c\rightarrow \eta_c) \;=\;
 {32 \alpha_s^2(2 m_c) |\Psi_{\eta_c}(0)|^2 \over 27  m_c^3}
\left( {773 \over 30} - 37 \ln 2 \right),
\end{equation}
\begin{equation}
P(\bar{b}\rightarrow B_c^+) \;=\; {8 \alpha_s^2(2 m_c)
|\Psi_{B_c}(0)|^2 \over 27  m_c^3}
    f\left({m_c \over m_b + m_c}\right),
\end{equation}
\begin{equation}
P(\bar{b}\rightarrow B_c^{*+}) \;=\; {8 \alpha_s^2(2 m_c)
|\Psi_{B_c^*}(0)|^2 \over 27  m_c^3}
     g\left({m_c \over m_b+m_c}\right),
\end{equation}
where the function $f(r)$ and  $g(r)$  are
\begin{eqnarray}
f(r) &=& {8 + 13r + 228r^2 - 212r^3 + 53r^4 \over 15 (1-r)^5} \nonumber\\
&&+{r
(1 + 8r + r^2 - 6r^3 + 2 r^4) \over (1-r)^6} \ln(r),~~~~
\end{eqnarray}
\begin{eqnarray}
g(r) &=& {24 + 109 r - 126 r^2 + 174 r^3 + 89 r^4 \over 15
(1-r)^5}\nonumber\\
&& + {r (7 - 4 r + 3 r^2 + 10 r^3 + 2 r^4) \over  (1-r)^6}
\ln(r).
\end{eqnarray}
$P_{b\to\Upsilon}$ and $P_{b\to\eta_b}$ can be obtained from (23)
and (24) by substituting the mass  $m_b$ for $m_c$. And $P_{c\to
B_c^{(*)+}}$ can be also got from (25) and (26) by interchanging
$m_b$ and $m_c$.

\begin{table*}
  \caption{The branching fractions for the quarkonium and $B_c$ meson production predicted by the complete leading order calculation and
  fragmentation approximation and their comparisons. }\label{tab:frag}
\begin{tabular*}{\textwidth}{@{\extracolsep{\fill}}cccc@{}}
  \hline  & {LO calculation ($B$)} &{fragmentation ($B^*)$}  & \quad {$\frac{B^*-B}{B}\quad$}\\\hline
   $\Gamma(W^+\rightarrow J/\psi c\bar{s})/ \Gamma(W^+\rightarrow c\bar{s})$ & $1.27\times10^{-4}$ & $1.31\times10^{-4}$  &$3.3\%$\\
     $\Gamma(W^+\rightarrow\eta_c c\bar{s})/ \Gamma(W^+\rightarrow c\bar{s})$ & $1.23\times10^{-4}$ & $1.27\times10^{-4}$  &$3.6\%$\\
     $\Gamma(W^+\rightarrow B_c^+ b\bar{s}) / \Gamma(W^+\rightarrow c\bar{s})$ & $8.84\times10^{-6}$ & $1.03\times10^{-5}$  &$17\%$\\
    $\Gamma(W^+\rightarrow B_c^{*+} b\bar{s})/ \Gamma(W^+\rightarrow c\bar{s})$ & $7.61\times10^{-6}$ & $8.92\times10^{-6}$  &$17\%$\\
    $\Gamma(W^+\rightarrow B_c^+ c\bar{c}) /\Gamma(W^+\rightarrow c\bar{b})$ & $3.64\times10^{-4}$ & $3.85\times10^{-4}$  &$6\%$\\
    $\Gamma(W^+\rightarrow B_c^{*+} c\bar{c}) / \Gamma(W^+\rightarrow c\bar{b})$ & $5.23\times10^{-4}$ & $5.41\times10^{-4}$  &$3\%$\\
    $\Gamma(W^+\rightarrow \eta_b c\bar{b}) / \Gamma(W^+\rightarrow c\bar{b})$ & $1.41\times10^{-5}$ & $1.76\times10^{-5}$  &$24\%$\\
    $\Gamma(W^+\rightarrow \Upsilon c\bar{b})/\Gamma(W^+\rightarrow c\bar{b})$ & $1.49\times10^{-5}$ & $1.81\times10^{-5}$  &$21\%$\\
  \hline
\end{tabular*}
\end{table*}

The numerical
results on the
  decay widths of $W$ inclusive decays to quarkonium and $B_c$ meson by using the fragmentation approximation
  are listed in Table \ref{tab:frag}. We also list the comparisons between the predictions given by the fragmentation approximation
  and the complete leading order calculations. We find that the difference
  between the predictions by the two methods is negligible for the charmonium productions. 
  However, for the bottomonium and $B_c^{(*)}$ productions, the differences is
sizable but understandable, since the power corrections to the fragmentation
mechanism is order of $2 m_b/m_w$, which is numerically around
$10\%$. 

\subsection{Theoretical uncertainties}
\begin{figure}[h]
\centering
\includegraphics[width=0.480\textwidth]{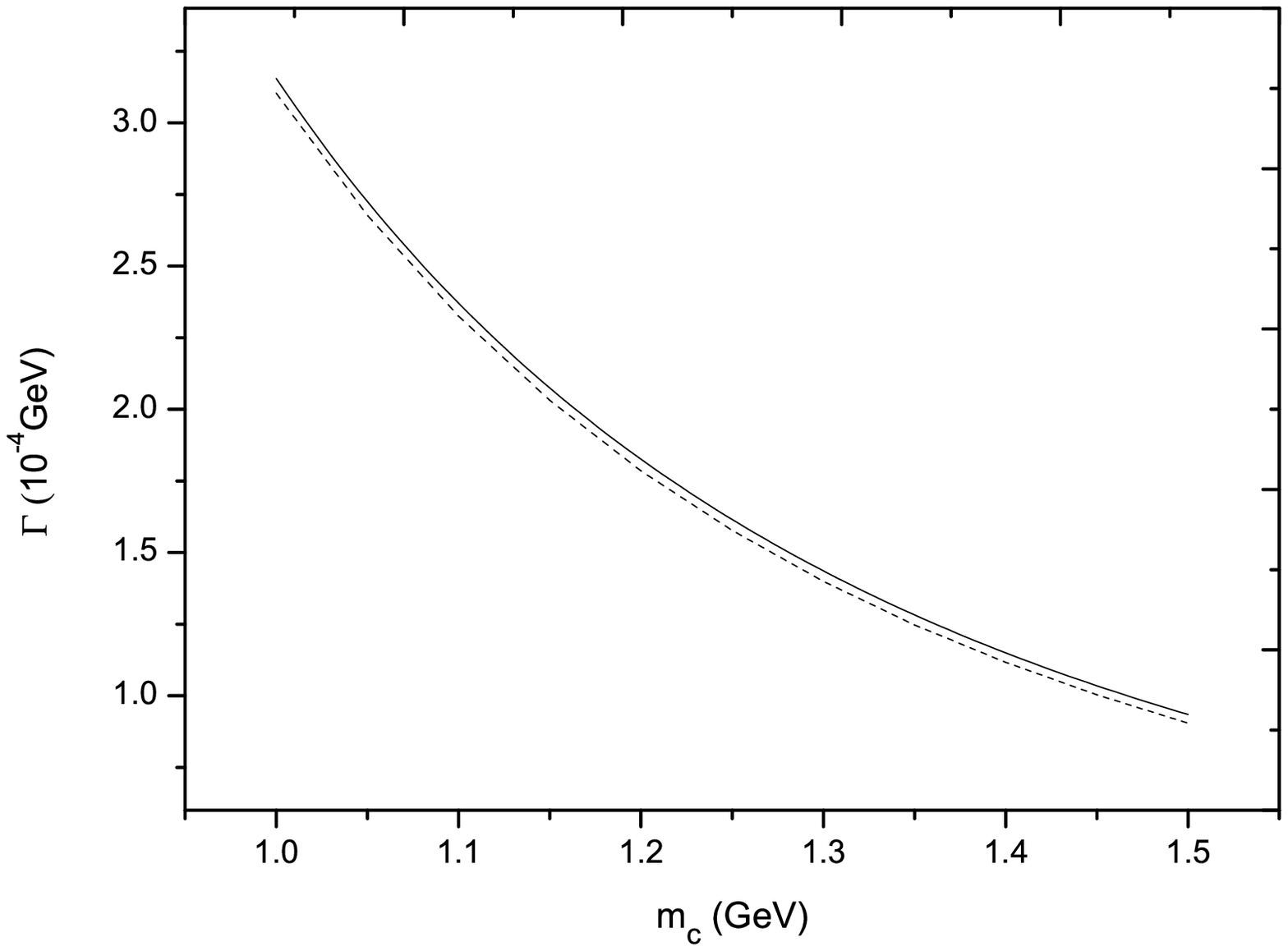}%
\hspace{5mm}
\includegraphics[width=0.480\textwidth]{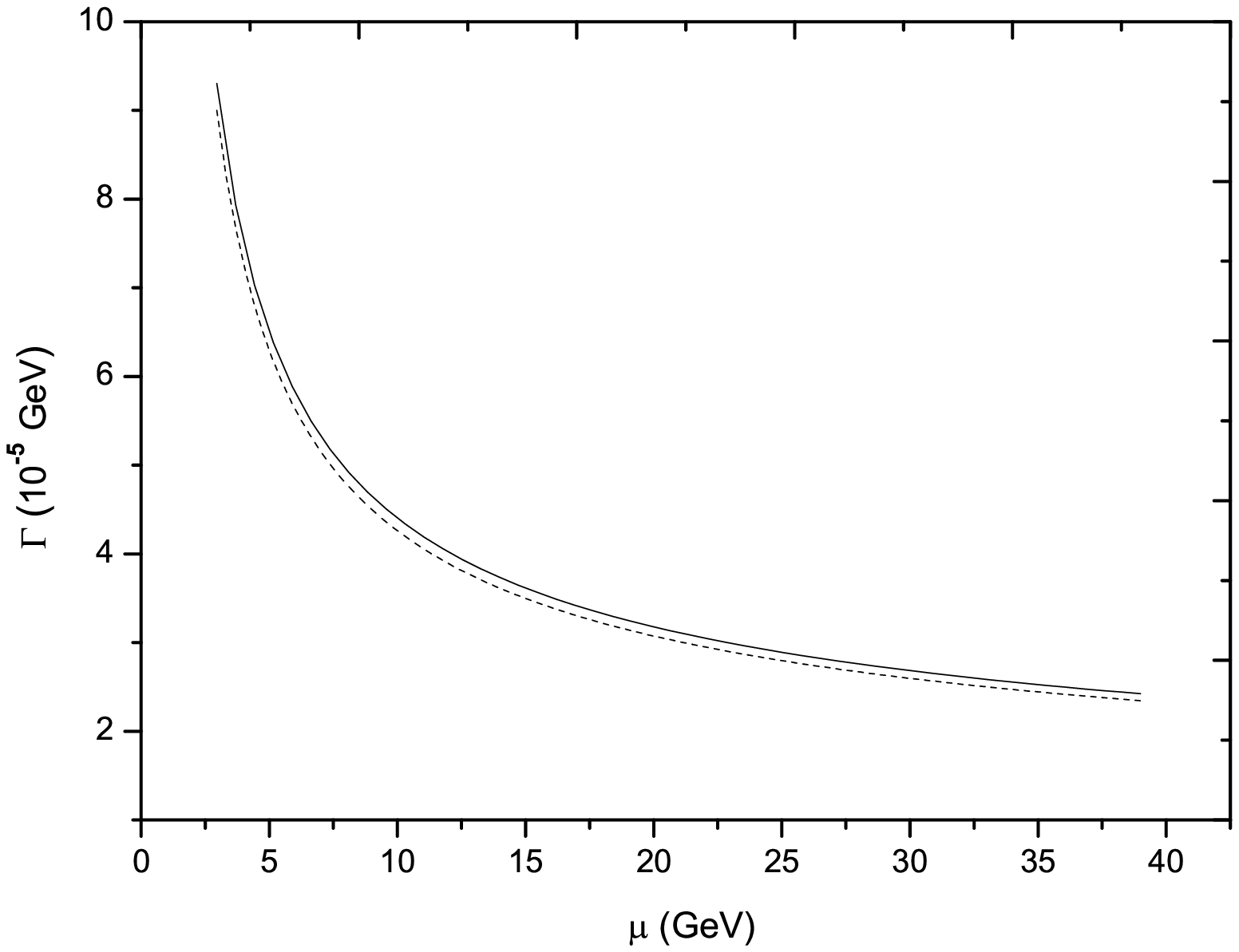}\hspace*{\fill}
\caption{\small The partial width of the process
$\Gamma(W^+\rightarrow
 J/\psi c\bar{s})$ versus the mass of charm quark(upper, $\mu=3~\mathrm{GeV}$)
 and the renormalization scale(lower, $m_c=1.5~\mathrm{GeV}$).
 The dashed line corresponds to the prediction
 by the complete LO calculation, while the solid line corresponds to
 that by the fragmentation approximation.} \label{fig:error} \vspace{-0mm}
\end{figure}

At last, we investigate the theoretical uncertainties in our
calculations for the $J/\psi$ and $\eta_c$ production from $W$ boson
decay. The main sources of uncertainties include the wave function
at the origin, the strong coupling constant and the quark
mass. After some numerical check, we find that the biggest
uncertainty in our results is from the choice of the charm quark
mass. Here we present the dependence of our results on the charm
quark mass in Fig.  \ref{fig:error}. It shows that we can get a production rate
of charmonium which is 1.6 times as much as we had calculated if we
adopt $m_c=1.27 ~\mathrm{GeV}$. Besides, the production rate from
fragmentation is a slightly bigger than that from complete leading
order computation. The difference between the LO calculation and the
fragmentation approximation is insensitive to the charm mass. The second
largest uncertainty is due to the choice of renormalization scale as mentioned
in the front of this section. Also we lay out their relations in Fig.  \ref{fig:error}.
The dependence on the renormalization scale can be reduced by considering the NLO QCD corrections and 
employing the Altarelli-Parisi equations to resum the large logarithms. 

\section{Conclusions}

In this paper, we show that there will be around $10^6$ $J/\Psi$ and
$\eta_c$, $10^4$ $B_c^{(*)}$ mesons and $10^3$ $\Upsilon$ and
$\eta_b$ annually produced from $W$ boson inclusive decay at the
LHC, if the machine works at the center energy
$\sqrt{s}=14\mathrm{TeV}$ and with the luminosity ${\cal
L}_{p-p}=10^{34} \mathrm{cm}^{-2}\mathrm{s}^{-1}$. And the
fragmentation mechanism works well for $W$ boson inclusive decays to
the S-wave charmonium, bottomonium and $B_c^{(*)}$ mesons. The
numerical calculation shows that the difference between
fragmentation approximation and complete LO calculation for $W$
boson decays to charmonium is around $3\%$, and it is not magnified
by considering the uncertainties of theoretical parameters. But for
$W$ decays to $B_c^{(*)}$ mesons and bottomonium,  the differences
rise, but it is understandable.

\section*{Acknowledgement}
 We thank Prof. Y.N. Gao, Y.J. Mao and Y.H. Zheng for the discussions on the detection of quarkonia at the LHC.
  This work is supported in part by the National Natural Science
 Foundation of China (NSFC) under the grants 10935012, 10928510,
 10821063 and 10705050, by the CAS Key Projects KJCX2-yw-N29 and
 H92A0200S2.

\appendix
\section{The square of the amplitudes for $W^+\rightarrow\eta_c+c+\bar{s}$}
Here we
introduce parameters
$$s_{10}=p_1\cdot p, \quad\quad s_{15}=p_1\cdot
p_5$$
with $p_1$, $p$ and $p_5$ being the momenta of $W^+$, $\eta_c$ and $\bar s$
quark, respectively.

\begin{eqnarray}
&&\Sigma|A_1|^2\nonumber\\&=&\frac{1}{(m_w^2(m_c^2 + m_w^2 - s_{10})^2(m_c^2 - m_w^2 + 2s_{15})^2)}\nonumber\\
&&\times(-64(12m_c^6m_w^2 - m_w^8 + 4s_{10}^2s_{15}(s_{10} + s_{15}) \nonumber\\&&+ 4m_w^6(2s_{10} + s_{15}) + 2m_w^2s_{10}(7s_{10}^2+ 4s_{15}^2  + 10s_{10}s_{15} ) \nonumber\\&&- m_w^4(19s_{10}^2 + 20s_{10}s_{15} + 4s_{15}^2) \nonumber\\&&+
   m_c^4(-5m_w^4  - 3s_{10}^2 + 24s_{10}s_{15} + 16s_{15}^2\nonumber\\&& + m_w^2(22s_{10} + 40s_{15})) - 2m_c^2(3m_w^6 - m_w^4(17s_{10} + 14s_{15})  \nonumber\\&&+ s_{10}(-s_{10}^2 + 10s_{10}s_{15} + 8s_{15}^2) \nonumber\\&&+
     m_w^2(21s_{10}^2 + 28s_{10}s_{15} + 8s_{15}^2))))\,,\nonumber
  \end{eqnarray}
  \begin{eqnarray}
  &&
 \Sigma|A_2|^2
 \nonumber\\&=&\frac{1}{(m_w^2(m_c^2 - m_w^2 + 2s_{15})^4)}\nonumber\\&&\times(64(27m_c^6m_w^2 + (m_w^2 - 2s_{15})^2(m_w^4 - 2m_w^2(s_{10} - 2s_{15}) \nonumber\\
&&- 4s_{15}(s_{10} + s_{15})) - m_c^2(m_w^2  - 2s_{15})(7m_w^4 \nonumber\\&&- 8s_{15}(5s_{10} + s_{15}) - 2m_w^2(10s_{10} + s_{15}))\nonumber\\&& -
   3m_c^4(7m_w^4  + 4(3s_{10} - 5s_{15})s_{15} + 2m_w^2(3s_{10} + 8s_{15}))))\, ,\nonumber
    \end{eqnarray}
    
  \begin{eqnarray}
   &&\Sigma {\rm Re}(A_1{A_2}^*) \nonumber\\&=&\frac{1}{(m_w^2(m_c^2 + m_w^2 - s_{10})(m_c^2 - m_w^2 + 2s_{15})^3)}
   \nonumber\\&&\times(64(9m_c^6m_w^2 + (m_w^2 - 2s_{15})(2m_w^6 + 4s_{10}s_{15}(s_{10} + s_{15}) \nonumber\\&&
   + 2m_w^2s_{10}(3s_{10} + 2s_{15})  - m_w^4(7s_{10} + 4s_{15})) \nonumber\\&&+ m_c^4(2s_{15}(21s_{10} + 4s_{15}) + 5m_w^2(3s_{10} + 8s_{15}))\nonumber\\&& +
   m_c^2(5m_w^6  - 16m_w^4s_{15} + m_w^2(-14s_{10}^2 \nonumber\\&&v+ 4s_{15}^2) + 4s_{15}(-5s_{10}^2 + 4s_{10}s_{15} + 4s_{15}^2))))\,.\nonumber
 \end{eqnarray}

\section{The square of the amplitudes for $W^+\rightarrow\Upsilon+c+\bar{b}$}
 Here we define parameters:
$${s_{10}}=p_1\cdot p, \quad\quad {s_{15}}=p_1\cdot p_5$$
with $p_1$, $p$ and $p_5$ being the momenta of $W^+$, $\Upsilon$ and $c$
quark, respectively.
\begin{eqnarray}&&
\Sigma|A_1|^2
\nonumber\\
&=&\frac{1}{(m_w^2(m_b^2 - m_c^2 + m_w^2 - {s_{10}})^2(-m_b^2 + m_c^2 + m_w^2 - 2{s_{15}})^2)}
\nonumber\\&&\times(64(66m_b^6m_w^2 - 2m_c^6m_w^2 + m_w^8 -
8m_w^6{s_{10}} + 35m_w^4{s^2_{10}} \nonumber\\&&- 46m_w^2{s^3_{10}}
- 4m_w^6{s_{15}}  + 20m_w^4{s_{10}}{s_{15}}
-20m_w^2{s^2_{10}}{s_{15}} \nonumber\\&&- 4{s^3_{10}}{s_{15}} + 4m_w^4{s^2_{15}}
  -8m_w^2{s_{10}}{s^2_{15}} - 4{s^2_{10}}{s^2_{15}} \nonumber\\&&-
m_c^4(3m_w^4 + m_w^2( 22{s_{10}} - 8{s_{15}}) - {s_{10}}(7{s_{10}} +
8{s_{15}}))\nonumber\\&&  +m_b^4(42m_c^2m_w^2 + m_w^4 + 3{s^2_{10}}
 - 32{s_{10}}{s_{15}} + 16{s^2_{15}} \nonumber\\&&- 2m_w^2(97{s_{10}} + 32{s_{15}})) -2m_c^2(m_w^4(15{s_{10}} \nonumber\\&& -
 2{s_{15}}) + m_w^2(-9{s^2_{10}} - 28{s_{10}}{s_{15}} + 4{s^2_{15}}) \nonumber\\&& + {s_{10}}(11{s^2_{10}} + 18{s_{10}}{s_{15}}
 + 8{s^2_{15}})) -
   2m_b^2(5m_c^4m_w^2 + 2m_w^6 \nonumber\\&&+ {s^2_{10}}({s_{10}} - 10{s_{15}}) + m_w^4(27{s_{10}} + 10{s_{15}})
    \nonumber\\&& - m_w^2(85{s^2_{10}} + 36{s_{10}}{s_{15}} + 4{s^2_{15}}) -
     m_c^2(25m_w^4 + 23{s^2_{10}}\nonumber\\&& + 28{s_{10}}{s_{15}} + 8{s^2_{15}} - 4m_w^2(7{s_{10}} + 5{s_{15}}))))) ,\nonumber
 \end{eqnarray}
 
  \begin{eqnarray}&&
 \Sigma|A_2|^2\nonumber\\&=&\frac{1}{(m_w^2(-m_b^2 + m_c^2 + m_w^2 - 2{s_{15}})^4)}
 \nonumber\\&&\times(64(27m_b^6m_w^2 - (m_c^2 + m_w^2 - 2{s_{15}})^2(m_c^2m_w^2 \nonumber\\&&
 - m_w^4 + 2m_w^2
 ({s_{10}} - 2{s_{15}})  + 4{s_{15}}({s_{10}} + {s_{15}})) \nonumber\\&&-
   m_b^2(m_c^2 + m_w^2 - 2{s_{15}})(3m_c^2m_w^2 + 7m_w^4 - 8{s_{15}}(5{s_{10}} + {s_{15}})\nonumber\\&&
   - 2m_w^2(10{s_{10}} + {s_{15}})) +
   m_b^4(73m_c^2m_w^2 - 21m_w^4 \nonumber\\&&- 6m_w^2(3{s_{10}} + 40{s_{15}})  + 4{s_{15}}(-9{s_{10}} +
   47{s_{15}})))),\nonumber
 \end{eqnarray}
 
  \begin{eqnarray}
  &&\Sigma {\rm Re}(A_1{A_2}^*)\nonumber\\&=&
  \frac{1}{(m_w^2(m_b^2 - m_c^2 + m_w^2 - {s_{10}})(-m_b^2 + m_c^2 + m_w^2 - 2{s_{15}})^3)}
  \nonumber\\&&\times(-64(-51m_b^6m_w^2 + m_b^4(-49m_c^2m_w^2 + 40m_w^4 \nonumber\\&&+ 2
  (13{s_{10}} - 44{s_{15}}){s_{15}}  + 7m_w^2(13{s_{10}} + 16{s_{15}})) \nonumber\\&&+
   m_b^2(3m_c^4m_w^2 + 25m_w^6 - 4m_w^4(7{s_{10}} + 6{s_{15}})\nonumber\\&&  - 2m_c^2(2m_w^4
   + 9{s^2_{10}} + 22m_w^2{s_{15}} - 2{s_{10}}{s_{15}} - 12{s^2_{15}})\nonumber\\&& +
     m_w^2(-38{s^2_{10}} - 24{s_{10}}{s_{15}} + 4{s^2_{15}})  \nonumber\\&&+ 4{s_{15}}(-{s^2_{10}} + 12{s_{10}}{s_{15}} + 4{s^2_{15}}))
     \nonumber\\&&+
   (m_c^2 + m_w^2 - 2{s_{15}})(m_c^4m_w^2 + 2m_w^6 + 4{s_{10}}{s_{15}}({s_{10}} + {s_{15}}) \nonumber\\&&
   + 2m_w^2{s_{10}}(3{s_{10}} + 2{s_{15}}) - m_w^4(7{s_{10}} + 4{s_{15}}) 
     \nonumber\\&&+
     m_c^2(3m_w^4 + 2{s_{10}}({s_{10}} + {s_{15}}) - m_w^2(3{s_{10}} + 2{s_{15}})))))
     .\nonumber
   \end{eqnarray}

\section{The square of the amplitudes for $B_c^{(*)}$ production }
First we present the squared matrix elements for $W^{+}\rightarrow
B_c^{+}+b+\bar{s}$. We introduce 
\[{s_{10}}=p_1\cdot p, \quad\quad {s_{15}}=p_1\cdot p_5\]
with $p_1$, $p$ and $p_5$ being the momenta of $W^{+}$, $B_c^{+}$ and $\bar{s}$
quark, respectively.
\begin{eqnarray}&&
\Sigma|A_1|^2\nonumber\\
&=&\frac{1}{
 m_b^2m_w^2(m_b(m_c^2 + m_w^2) + m_c(m_c^2 + m_w^2 - 2s_{10}))^2}
\nonumber\\&&\times\frac{1}{(m_c^2 - m_w^2 + 2s_{15})^2}
(16(m_b + m_c)^2(2m_b^5m_cm_w^2(m_c^2 \nonumber\\&&
- m_w^2 - 2s_{15}) 
+ m_b^4(m_c^4m_w^2 + m_w^2(-m_w^4 + 2m_w^2(s_{10} - 2s_{15})\nonumber\\&& + 8s_{15}(s_{10} + s_{15})) - 2m_c^2(2s_{15}(s_{10} + s_{15}) 
 \nonumber\\&&+ 3m_w^2(3s_{10} + 4s_{15})))-
   m_c^2(m_c^2 + m_w^2 - 2s_{10})(3m_c^4m_w^2 \nonumber\\&&
   - m_w^6 + 4m_w^4(s_{10} + s_{15}) 
   - 8s_{10}s_{15}(s_{10} + s_{15}) \nonumber\\&&- 4m_w^2(s_{10}^2 + s_{10}s_{15} + s_{15}^2) +
     m_c^2(-2m_w^4 + 4m_w^2(s_{10} + s_{15})\nonumber\\&& + 4(-s_{10}^2 + s_{10}s_{15} + s_{15}^2))) 
     - 2m_bm_c(7m_c^6m_w^2\nonumber\\&& + m_c^4(-m_w^4  - 4s_{10}^2 + 12s_{10}s_{15} + 8s_{15}^2
      + 2m_w^2(s_{10} + 6s_{15}))\nonumber\\&& -
     m_w^2(m_w^2 - 2s_{10})(m_w^4 - 2m_w^2(3s_{10} + 2s_{15}) 
    \nonumber\\&& + 4(2s_{10}^2 + 3s_{10}s_{15} + s_{15}^2))
 - m_c^2(5m_w^6 + 8s_{10}s_{15}(3s_{10} + 2s_{15})   \nonumber\\&&
   - 2m_w^4(11s_{10} + 8s_{15}) + 4m_w^2(6s_{10}^2 + 6s_{10}s_{15} + s_{15}^2))) \nonumber\\&&      + m_b^2(-22m_c^6m_w^2 + m_c^4(11m_w^4 - 8s_{15}(5s_{10} + 3s_{15})  \nonumber\\&&
  - 4m_w^2(7s_{10} + 13s_{15}))       + 2m_c^2(5m_w^6 + 8s_{10}s_{15}(s_{10} + s_{15}) \nonumber\\&&
- m_w^4(29s_{10} + 22s_{15}) 
      + 12m_w^2(3s_{10}^2 + 4s_{10}s_{15} + s_{15}^2))\nonumber\\&&
 +m_w^2(m_w^6 - 16s_{10}(s_{10} + s_{15})^2 
      - 2m_w^4(5s_{10} + 2s_{15})\nonumber\\&& + 4m_w^2(6s_{10}^2 + 7s_{10}s_{15} + s_{15}^2)))
       \nonumber\\&&
      -2m_b^3m_c(6m_c^4m_w^2 + m_c^2(-5m_w^4 + 4s_{15}(3s_{10} + 2s_{15}) \nonumber\\&&
      + m_w^2(20s_{10} + 26s_{15})) - m_w^2(m_w^4 - 12m_w^2(s_{10} + s_{15}) \nonumber\\&&
      + 4(4s_{10}^2 + 8s_{10}s_{15} + 3s_{15}^2))))),\nonumber
  \end{eqnarray}

  \begin{eqnarray}&&
   \Sigma|A_2|^2\nonumber\\
   &=&\frac{1}{(m_b^2m_w^2(m_c^2 - m_w^2 + 2s_{15})^4)}
   \nonumber\\&&\times
   (16(m_b + m_c)^2(2m_b^3m_c(m_c^2m_w^2 - m_w^4 \nonumber\\&&- 10m_w^2s_{15} + 8s_{15}^2) + 2m_bm_c(6m_c^4m_w^2 - (m_w^2 - 2s_{15})(m_w^4 \nonumber\\&&- 8s_{10}s_{15} - 2m_w^2(2s_{10} + s_{15})) -
     m_c^2(5m_w^4 + 8(s_{10} - s_{15})s_{15} \nonumber\\&&+ 2m_w^2(2s_{10} + s_{15}))) + (m_c^2 - m_w^2 + 2s_{15})(4m_c^4m_w^2 \nonumber\\&&- (m_w^2 - 2s_{15})(m_w^4 - 2m_w^2(s_{10} - 2s_{15}) - 4s_{15}(s_{10} + s_{15}))\nonumber\\&& +
     m_c^2(3m_w^4 - 4s_{15}(4s_{10} + 3s_{15}) + m_w^2(-8s_{10} + 12s_{15}))) \nonumber\\&& + m_b^2(9m_c^4m_w^2 - (m_w^2 - 2s_{15})(m_w^4 - 4(s_{10} - 2s_{15})s_{15} \nonumber\\&& - 2m_w^2(s_{10} + 7s_{15})) -
     2m_c^2(4m_w^4 + 2(s_{10} - 10s_{15})s_{15}\nonumber\\&& + m_w^2(s_{10} + 22s_{15}))))),\nonumber
   \end{eqnarray}
   
  \begin{eqnarray}
   &&\Sigma {\rm Re}(A_1{A_2}^*)\nonumber\\
   &=&\frac{1}{m_b^2m_w^2(m_b(m_c^2 + m_w^2) + m_c(m_c^2 + m_w^2 - 2s_{10}))}
   \nonumber\\&&\times\frac{1}{(m_c^2 - m_w^2 + 2s_{15})^3}(-16(m_b + m_c)^2(2m_b^4m_c(m_c^2m_w^2 \nonumber\\&&
   - 3m_w^4 - 2m_w^2s_{15} + 2s_{15}^2) - m_b^3(3m_c^4m_w^2  \nonumber\\&&+ 2m_c^2(3m_w^4 + 2(s_{10} - 2s_{15})s_{15} + 5m_w^2(s_{10} + 3s_{15})) \nonumber\\&&
   +m_w^2(3m_w^4 + 4s_{15}(s_{10} + 2s_{15}) 
   - 2m_w^2(3s_{10} + 5s_{15}))) \nonumber\\&&
   - 4m_b^2m_c(3m_c^4m_w^2 + 2m_w^6 + 2s_{15}^2(2s_{10} + s_{15}) \nonumber\\&&
   - 2m_w^4(s_{10} + 3s_{15}) + m_w^2(-2s_{10}^2 + 3s_{15}^2) \nonumber\\&&
   +m_c^2(-2m_w^4 + s_{15}(7s_{10} + s_{15}) + m_w^2(4s_{10} + 11s_{15}))) \nonumber\\&&
   + m_c(-((m_w^2 - 2s_{10})(m_w^4 - 4s_{15}^2)(m_w^2 - 2(s_{10} + s_{15}))) \nonumber\\&&+m_c^4(m_w^4 - 2m_w^2(s_{10} - s_{15})  - 4s_{15}(3s_{10} + 2s_{15}))\nonumber\\&&
   - 2m_c^2(m_w^4(s_{10} - 2s_{15}) + 4s_{15}(-3s_{10}^2 - s_{10}s_{15} + s_{15}^2) \nonumber\\&&
   + m_w^2(-2s_{10}^2 + 8s_{10}s_{15} + 2s_{15}^2))) +
   m_b(-5m_c^6m_w^2 \nonumber\\&&
   + m_c^4(3m_w^4 - 2m_w^2(s_{10} + 2s_{15}) - 8s_{15}(5s_{10} + 2s_{15})) \nonumber\\&&
   - m_w^2(m_w^2 - 2s_{15})(3m_w^4 -  2m_w^2(5s_{10} + 4s_{15})\nonumber\\&& 
   + 4(2s_{10}^2 + 3s_{10}s_{15} + s_{15}^2)) +m_c^2(m_w^6 - 6m_w^4(2s_{10} + s_{15})\nonumber\\&&
    - 8s_{15}(-2s_{10}^2 + 3s_{10}s_{15} + 2s_{15}^2) + 4m_w^2(4s_{10}^2 \nonumber\\&&
    + 5s_{10}s_{15} + 4s_{15}^2))))).\nonumber
   \end{eqnarray}

 For $W^{+}\rightarrow
B_c^{*+}+b+\bar{s}$, we introduce 
$${s_{10}}=p_1\cdot p, \quad\quad {s_{15}}=p_1\cdot p_5$$
with $p_1$, $p$ and $p_5$ being the momenta of $W^{+}$, $B_c^{*+}$ and $\bar{s}$
quark, respectively.
 \begin{eqnarray}
&&
\Sigma|A_1|^2\nonumber\\
&=&\frac{1}{
 m_b^2m_w^2(m_b(m_c^2 + m_w^2) + m_c(m_c^2 + m_w^2 - 2s_{10}))^2}
\nonumber\\&&\times\frac{1}{(m_c^2 - m_w^2 + 2s_{15})^2}
(16(m_b + m_c)^2(6m_b^5m_cm_w^2(m_c^2 - m_w^2 \nonumber\\&&
- 2s_{15}) + 2m_b^3m_c(40m_c^4m_w^2 - m_w^2(3m_w^4 - 48s_{10}^2 - 48s_{10}s_{15} \nonumber\\&&
+ 4s_{15}^2 + 4m_w^2(3s_{10} + s_{15})) -m_c^2(5m_w^4 + 4(5s_{10} - 2s_{15})s_{15}\nonumber\\&&
 + 2m_w^2(50s_{10} + 23s_{15}))) + m_b^4(35m_c^4m_w^2 + m_w^2(-3m_w^4 \nonumber\\&&
 + 8(3s_{10} - s_{15})s_{15} + m_w^2(6s_{10} + 4s_{15})) -
 2m_c^2(8m_w^4 \nonumber\\&&+ 2(3s_{10} - s_{15})s_{15} + m_w^2(27s_{10} + 28s_{15}))) + m_c^2(m_c^2 + m_w^2 \nonumber\\&&
 - 2s_{10})(9m_c^4m_w^2 + m_w^6 - 4m_w^4(s_{10} + s_{15}) + 8s_{10}s_{15}(s_{10} + s_{15}) \nonumber\\&&
 +4m_w^2(5s_{10}^2 + s_{10}s_{15} + s_{15}^2) - 2m_c^2(m_w^4 + 2m_w^2(7s_{10} + s_{15})\nonumber\\&&
  - 2(s_{10}^2 - s_{10}s_{15} + s_{15}^2))) +2m_bm_c(23m_c^6m_w^2 
  + m_c^2(-m_w^6 \nonumber\\&&
  + 24s_{10}^2s_{15} - 2m_w^4(21s_{10} + 8s_{15}) 
  + 12m_w^2(10s_{10}^2 + 2s_{10}s_{15}\nonumber\\&&
   + s_{15}^2))
   +m_w^2(m_w^2 - 2s_{10})(m_w^4 
   - 2m_w^2(3s_{10} + 2s_{15})\nonumber\\&&
    + 4(6s_{10}^2 + 3s_{10}s_{15} + s_{15}^2)) 
   + m_c^4(9m_w^4 - 2m_w^2(47s_{10} + 6s_{15}) \nonumber\\&&
   + 4(s_{10}^2 - 3s_{10}s_{15} + 2s_{15}^2))) 
   +m_b^2(88m_c^6m_w^2 + m_c^4(11m_w^4 \nonumber\\&&
   + 24s_{15}(-2s_{10} + s_{15}) 
   - 4m_w^2(72s_{10}17s_{15})) + m_w^2(m_w^6 \nonumber\\&&- 2m_w^4(5s_{10} + 2s_{15})
    - 16s_{10}(3s_{10}^2 + 2s_{10}s_{15} + s_{15}^2)\nonumber\\&& +4m_w^2(10s_{10}^2 + 7s_{10}s_{15} + s_{15}^2)) - 2m_c^2(2m_w^6 - 8s_{10}^2s_{15} \nonumber\\&&+ m_w^4(43s_{10} + 18s_{15}) - 4m_w^2(33s_{10}^2 + 14s_{10}s_{15} + 2s_{15}^2))))),\nonumber
  \end{eqnarray}
  
  \begin{eqnarray}&&
   \Sigma|A_2|^2\nonumber\\
   &=&\frac{1}{(m_b^2m_w^2(m_c^2 - m_w^2 + 2s_{15})^4)}
(16(m_b + m_c)^2(6m_b^3m_c(m_c^2m_w^2 \nonumber\\&&- m_w^4 - 10m_w^2s_{15} + 8s_{15}^2) + (m_c^2 - m_w^2 + 2s_{15})(2m_c^4m_w^2 \nonumber\\&&- (m_w^2 - 2s_{15})(m_w^4 - 2m_w^2(s_{10} - 2s_{15}) - 4s_{15}(s_{10} + s_{15})) \nonumber\\&&+m_c^2(m_w^4 + 4s_{15}(-2s_{10} + s_{15}) - 4m_w^2(s_{10} + 2s_{15})))\nonumber\\&&
 + 2m_bm_c(4m_c^4m_w^2 - (m_w^2 - 2s_{15})(m_w^4 - 8s_{10}s_{15} - 2m_w^2(2s_{10} \nonumber\\&&+ s_{15})) -m_c^2(3m_w^4 + 8(s_{10} - 3s_{15})s_{15} + m_w^2(4s_{10} + 30s_{15}))) \nonumber\\&&
 + m_b^2(11m_c^4m_w^2 - 3m_w^6 + 6m_w^4s_{10}  + 20m_w^2s_{15}^2- 8s_{15}^2(3s_{10} \nonumber\\&&
 + 2s_{15}) - 2m_c^2(4m_w^4 + 2(3s_{10} - 22s_{15})s_{15} \nonumber\\&&+ m_w^2(3s_{10} + 58s_{15}))))),\nonumber
   \end{eqnarray}
   
  \begin{eqnarray}&&\Sigma {\rm Re}(A_1{A_2}^*)\nonumber\\
   &=&\frac{1}{m_b^2m_w^2(m_b(m_c^2 + m_w^2) + m_c(m_c^2 + m_w^2 - 2s_{10}))}
   \nonumber\\&&\times\frac{1}{(m_c^2 - m_w^2 + 2s_{15})^3}(-16(m_b + m_c)^2(6m_b^4m_c(m_c^2m_w^2 - 3m_w^4\nonumber\\&&
    - 2m_w^2s_{15} + 2s_{15}^2) + m_b^3(31m_c^4m_w^2 - 2m_c^2(25m_w^4 
    + 6(s_{10} \nonumber\\&&
    - 4s_{15})s_{15} + 3m_w^2(5s_{10} + 7s_{15})) +m_w^2(-9m_w^4 
    - 4s_{15}(3s_{10} \nonumber\\&&+ 2s_{15}) + 2m_w^2(9s_{10} + 7s_{15})))     + 4m_b^2m_c(10m_c^4m_w^2 \nonumber\\&&
- 7m_w^6 - 2s_{15}^2(2s_{10} + s_{15}) 
+ 2m_w^4(5s_{10} + 6s_{15}) \nonumber\\&& +m_w^2(6s_{10}^2 - 4s_{10}s_{15}- 7s_{15}^2) - m_c^2(6m_w^4 + (5s_{10} - 17s_{15})s_{15} \nonumber\\&&
+ m_w^2(20s_{10} + 23s_{15})))+m_c(4m_c^6m_w^2 - (m_w^2 - 2s_{10})(m_w^4 \nonumber\\&&
 - 4s_{15}^2)(m_w^2 - 2(s_{10} + s_{15})) 
+ m_c^4(5m_w^4 - 4(s_{10} - 2s_{15})s_{15} \nonumber\\&&
- 14m_w^2(s_{10} + s_{15})) - 2m_c^2(-6m_w^2(s_{10} + s_{15})^2 \nonumber\\&&+ 3m_w^4(s_{10} + 2s_{15}) + 4s_{15}(-s_{10}^2 + 3s_{10}s_{15} + s_{15}^2))) \nonumber\\&&+m_b(21m_c^6m_w^2 
+ m_c^4(7m_w^4 + 8s_{15}(-2s_{10} + 5s_{15})\nonumber\\&& - 2m_w^2(29s_{10}+ 32s_{15}))  
- m_w^2(m_w^2 - 2s_{15})(3m_w^4 \nonumber\\&&
- 2m_w^2(5s_{10} + 4s_{15})
 + 4(2s_{10}^2 + 3s_{10}s_{15} + s_{15}^2)) \nonumber\\&&+m_c^2(-13m_w^6 + m_w^4(4s_{10} - 2s_{15}) 
 - 8s_{15}^2(7s_{10} + 2s_{15})\nonumber\\&& + 4m_w^2(10s_{10}^2 + 13s_{10}s_{15}+ 4s_{15}^2))))).\nonumber
   \end{eqnarray}

Then, we present the squared matrix elements for
$W^{+}\rightarrow B_c^{+}+c+\bar{c}$. Here we introduce 
$${s_{10}}=p_1\cdot p, \quad\quad {s_{15}}=p_1\cdot p_5$$
with $p_1$, $p$ and $p_5$ being the momenta of $W^{+}$, $B_c^{+}$ and $\bar{c}$
quark, respectively.
 \begin{eqnarray}&&
 \Sigma|A_1|^2\nonumber\\
 &=&\frac{1}{(m_b^3
    + m_b^2m_c - m_c^3 + m_cm_w^2 + m_b(-m_c^2 + m_w^2 - 2{s_{10}}))^2}
 \nonumber\\&&\times 
 \frac{1}{m_c^2m_w^2(m_b^2 - m_c^2 + m_w^2
    - 2{s_{10}} - 2{s_{15}})^2}
 \nonumber\\&&\times(-16(m_b + m_c)^2(3m_b^8m_w^2 + 14m_b^7m_cm_w^2 \nonumber\\&&- 2m_b^5m_c(9m_c^2m_w^2 - 19m_w^4
  + 8{s^2_{10}} - 4{s_{10}}{s_{15}} - 8{s^2_{15}} \nonumber\\&&
  +14m_w^2({s_{10}} + 2{s_{15}})) + m_b^6(14m_c^2m_w^2
 + 9m_w^4 \nonumber\\&&- 2m_w^2(5{s_{10}} + 6{s_{15}}) + 4(-{s^2_{10}} + {s_{10}}{s_{15}} + {s^2_{15}})) \nonumber
 \\&&-2m_b^3m_c(19m_c^4m_w^2 - 7m_w^6 + m_w^4(26{s_{10}} + 8{s_{15}}) \nonumber\\&&- 4m_w^2({s^2_{10}} + 12{s_{10}}{s_{15}} - {s^2_{15}}) +8{s_{10}}(-{s^2_{10}} + {s_{10}}{s_{15}} + 2{s^2_{15}}) \nonumber
 \\&&- 2m_c^2(9m_w^4 - 2{s^2_{10}} +
 2m_w^2({s_{10}} - 10{s_{15}}) + 4{s_{10}}{s_{15}} \nonumber\\&&+ 8{s^2_{15}})) -m_b^2(30m_c^6m_w^2 - m_c^4(23m_w^4 -
  12{s^2_{10}} \nonumber\\&&+ 2m_w^2({s_{10}} - 34{s_{15}}) + 44{s_{10}}{s_{15}} + 28{s^2_{15}}) \nonumber\\&&+(m_w^2 - 2{s_{10}})(m_w^2
   - 2{s_{15}})(m_w^4 + m_w^2(4{s_{10}} - 2{s_{15}}) \nonumber\\&&- 4{s_{10}}({s_{10}} + {s_{15}})) -4m_c^2(3m_w^6 -
   12m_w^4{s_{10}} \nonumber\\&&- 4{s_{10}}{s_{15}}({s_{10}} + {s_{15}}) + m_w^2(3{s^2_{10}} + 20{s_{10}}{s_{15}} - 6{s^2_{15}})))
   \nonumber\\&&+m_c^2(-7m_c^6m_w^2 - m_w^2(m_w^2 - 2{s_{15}})(m_w^4 + 8{s_{10}}{s_{15}}\nonumber\\&& - 2m_w^2({s_{10}} + {s_{15}}))
   +m_c^2(-m_w^6 + 8{s^3_{10}} - 4m_w^4({s_{10}} - 2{s_{15}}) \nonumber\\&&+ 4m_w^2({s^2_{10}} + 2{s_{10}}{s_{15}} - 2{s^2_{15}}))
    +m_c^4(9m_w^4 \nonumber\\&&- 2m_w^2(3{s_{10}} + 10{s_{15}}) + 4(3{s^2_{10}} + 3{s_{10}}{s_{15}} + {s^2_{15}})))
    \nonumber\\&&-2m_bm_c(11m_c^6m_w^2 + m_w^2(m_w^2 - 2{s_{10}})(m_w^2 - 2{s_{15}})(m_w^2 \nonumber\\&&+ 2{s_{10}} - 2{s_{15}})
    -m_c^2(m_w^6 - 8{s^2_{10}}{s_{15}} + m_w^4(-10{s_{10}} + 8{s_{15}})  \nonumber\\&&
    + 4m_w^2(2{s^2_{10}} + 4{s_{10}}{s_{15}}
    - 3{s^2_{15}})) -m_c^4(11m_w^4 \nonumber\\&&- 2m_w^2(3{s_{10}} + 14{s_{15}})
     + 4({s^2_{10}} + 5{s_{10}}{s_{15}}
   + 2{s^2_{15}}))) \nonumber\\&& +m_b^4(-44m_c^4m_w^2 + 5m_w^6 - 4m_w^4(5{s_{10}} + 2{s_{15}}) \nonumber\\
    &&+ 8m_w^2{s_{10}}({s_{10}} + 5{s_{15}}) + 8{s_{10}}({s^2_{10}} - 2{s_{10}}{s_{15}} - 2{s^2_{15}}) \nonumber\\&&+m_c^2(55m_w^4
    - 2m_w^2(9{s_{10}} + 46{s_{15}})\nonumber\\&& + 4(-3{s^2_{10}} + {s_{10}}{s_{15}} + 7{s^2_{15}}))))),\nonumber
  \end{eqnarray}

  \begin{eqnarray}&&
   \Sigma|A_2|^2\nonumber\\
   &=&\frac{1}{m_c^2m_w^2(m_b^2 - m_c^2 + m_w^2 - 2{s_{10}} - 2{s_{15}})^4}  
   \nonumber\\&&\times(16(m_b + m_c)^2(4m_b^6m_w^2 + 12m_b^5m_cm_w^2 - 6m_c^6m_w^2 \nonumber\\&&+ (m_w^2 - 2({s_{10}} +
   {s_{15}}))^2(m_w^4 - 2m_w^2({s_{10}} - 2{s_{15}}) \nonumber\\&&- 4{s_{15}}({s_{10}} + {s_{15}})) - m_c^4(-13m_w^4 +
   8m_w^2({s_{10}} + {s_{15}})\nonumber\\&& + 12({s_{10}} + {s_{15}})^2) +
   m_b^4(2m_c^2m_w^2 + 7m_w^4 \nonumber\\&&+ 4m_w^2(-5{s_{10}} + {s_{15}}) + 4({s^2_{10}} - 2{s_{10}}{s_{15}} -
   3{s^2_{15}})) \nonumber\\&&+ 2m_b^3m_c(m_w^4 - 2m_w^2(13{s_{10}} + 7{s_{15}}) + 8(2{s^2_{10}} + 3{s_{10}}{s_{15}}\nonumber\\&&
   + {s^2_{15}})) -2m_c^2(4m_w^6 + 4({s_{10}} + {s_{15}})^2(3{s_{10}} + 4{s_{15}}) \nonumber\\&&- m_w^4(7{s_{10}} +
   4{s_{15}}) - 8m_w^2({s^2_{10}} + 3{s_{10}}{s_{15}} + 2{s^2_{15}})) \nonumber\\&& 
   +2m_bm_c(2m_c^4m_w^2- m_w^6 -
   16{s_{10}}({s_{10}} + {s_{15}})^2 \nonumber\\&& + m_w^4(-8{s_{10}} + 4{s_{15}}) + 4m_w^2(7{s^2_{10}} + 6{s_{10}}{s_{15}} -
   {s^2_{15}}) \nonumber\\&&+
   m_c^2(-m_w^4 + 2m_w^2({s_{10}} - 5{s_{15}}) + 8{s_{15}}({s_{10}} + {s_{15}}))) \nonumber\\&&+
   2m_b^2(8m_c^4m_w^2 + 2m_w^6 - 4({s_{10}} - 4{s_{15}})({s_{10}} + {s_{15}})^2 \nonumber\\&&
   + m_w^4
   (-11{s_{10}} + 4{s_{15}}) +8m_w^2(2{s^2_{10}} - {s_{10}}{s_{15}} - 3{s^2_{15}}) \nonumber\\&&
   - 2m_c^2(5m_w^4 +
   m_w^2(5{s_{10}} + 11{s_{15}}) \nonumber\\&&- 2(5{s^2_{10}} + 12{s_{10}}{s_{15}} + 7{s^2_{15}}))))),\nonumber
  \end{eqnarray}
  
  \begin{eqnarray}&&
   \Sigma {\rm Re}(A_1{A_2}^*)\nonumber\\
   &=&\frac{1}{m_b^3 + m_b^2m_c - m_c^3 + m_cm_w^2 + m_b(-m_c^2 + m_w^2 - 2{s_{10}})}\nonumber\\
 &&\times\frac{1}{m_c^2m_w^2(m_b^2 - m_c^2 + m_w^2 - 2{s_{10}} - 2{s_{15}})^3}
   \nonumber\\&&\times(16(m_b + m_c)^2(5m_b^6m_cm_w^2 + m_b^5(10m_c^2m_w^2 + 5m_w^4 \nonumber\\&&-
   4{s^2_{10}} - 14m_w^2{s_{15}} + 4{s_{10}}{s_{15}} + 8{s^2_{15}}) +m_b^4m_c(-17m_c^2m_w^2 \nonumber\\&&
   +17m_w^4 + 6m_w^2({s_{10}} - 6{s_{15}}) - 8(3{s^2_{10}} + {s_{10}}{s_{15}} - 2{s^2_{15}}))
   \nonumber\\&&
   +m_b^2m_c(-29m_c^4m_w^2 + 5m_w^6 - 2m_w^4(3{s_{10}} + 11{s_{15}}) \nonumber\\&&- 4m_w^2(7{s^2_{10}}
    - 9{s_{10}}{s_{15}} - 8{s^2_{15}}) +8(3{s^3_{10}} + 2{s^2_{10}}{s_{15}} 
    \nonumber\\&&
    - 3{s_{10}}{s^2_{15}} - 2{s^3_{15}})
    + 2m_c^2(9m_w^4 - 2({s_{10}} + 2{s_{15}})^2 \nonumber\\&&
    + m_w^2(3{s_{10}} + 4{s_{15}}))) -
   2m_b^3(26m_c^4m_w^2 - 4m_w^6 \nonumber\\&&
   - 14m_w^2{s_{15}}(2{s_{10}} + {s_{15}})+ m_w^4(9{s_{10}} +
   14{s_{15}}) \nonumber\\&&+
   4(-{s^3_{10}} 
   + 2{s^2_{10}}{s_{15}} + 4{s_{10}}{s^2_{15}} + {s^3_{15}})
   + m_c^2(-15m_w^4  \nonumber\\&&+
   2{s_{10}}(5{s_{10}} + 6{s_{15}}) + m_w^2({s_{10}} + 18{s_{15}}))) \nonumber\\&&+m_c(9m_c^6m_w^2 + m_w^2
   (m_w^2 - 2{s_{15}})(m_w^4 - 4m_w^2{s_{10}} + 4{s^2_{10}} - 4{s^2_{15}}) \nonumber\\&&
   + m_c^4(-11m_w^4 + 28m_w^2({s_{10}} + {s_{15}}) - 4{s_{10}}({s_{10}} + 2{s_{15}}))\nonumber\\&&
    +m_c^2(m_w^6
   - 8{s_{10}}{s_{15}}({s_{10}} + {s_{15}}) - 2m_w^4(6{s_{10}} + 7{s_{15}}) \nonumber\\&&+ 4m_w^2(3{s^2_{10}} +
   8{s_{10}}{s_{15}} + 6{s^2_{15}}))) +
   m_b(10m_c^6m_w^2 \nonumber\\&&+ (m_w^2 - 2{s_{10}})(m_w^2 - 2{s_{15}})(3m_w^4 - 8m_w^2({s_{10}} + {s_{15}})
   \nonumber\\&&+ 4({s_{10}} + {s_{15}})^2) -m_c^4(11m_w^4 + 8{s^2_{10}} + 12{s_{10}}{s_{15}} + 8{s^2_{15}}
   \nonumber\\&&- 2m_w^2(21{s_{10}} + 25{s_{15}})) -2m_c^2(m_w^6 + 4m_w^4{s_{15}} \nonumber\\&&+
   2m_w^2(3{s^2_{10}} - 3{s_{10}}{s_{15}} - 7{s^2_{15}}) - 4({s^3_{10}} + 2{s^2_{10}}{s_{15}} - {s^3_{15}}))))).\nonumber
  \end{eqnarray}

   For $W^{+}\rightarrow B_c^{*+}+c+\bar{c}$, we introduce parameters:
$${s_{10}}=p_1\cdot p, \quad\quad {s_{15}}=p_1\cdot p_5$$
with $p_1$, $p$ and $p_5$ being the momenta of $W^{+}$, $B_c^{*+}$ and $\bar{c}$
quark, respectively.
 \begin{eqnarray}
 &&
 \Sigma|A_1|^2\nonumber\\
 &=&\frac{1}{(m_b^3
    + m_b^2m_c - m_c^3 + m_cm_w^2 + m_b(-m_c^2 + m_w^2 - 2{s_{10}}))^2}
 \nonumber\\&&\times 
 \frac{1}{m_c^2m_w^2(m_b^2 - m_c^2 + m_w^2
    - 2{s_{10}} - 2{s_{15}})^2}
 \nonumber\\&&\times(16(m_b + m_c)^2(9m_b^8m_w^2 + 46m_b^7m_cm_w^2 + m_b^6(74m_c^2m_w^2 \nonumber\\&&
 + 7m_w^4 - 2m_w^2(27s_{10} + 2s_{15}) + 4(3s_{10}^2 + 3s_{10}s_{15} + s_{15}^2)) \nonumber\\&&
 + 2m_b^5m_c(27m_c^2m_w^2 + 5m_w^4 - 2m_w^2(55s_{10} + 2s_{15}) \nonumber\\&&
 + 4(6s_{10}^2 + 7s_{10}s_{15} + 2s_{15}^2))+m_b^4(68m_c^4m_w^2 - m_w^6 \nonumber\\&&
 - 8s_{10}^2(3s_{10} + 2s_{15}) 
 - 4m_w^4(7s_{10} + 2s_{15}) \nonumber\\&&
 + 8m_w^2 (12s_{10}^2 + s_{10}s_{15} + s_{15}^2) 
 +m_c^2(-39m_w^4 + 60s_{10}^2 \nonumber\\&&
 + 92s_{10}s_{15} + 28s_{15}^2 + m_w^2(-254s_{10} + 44s_{15})))\nonumber\\&&
  -2m_bm_c(7m_c^6m_w^2 - m_w^2(m_w^2 - 2s_{10})(m_w^4 + 2m_w^2(s_{10} \nonumber\\&&
  - 2s_{15})  + 4(4s_{10}^2 - s_{10}s_{15} + s_{15}^2))
  + m_c^4(-21m_w^4 \nonumber\\&&
  + 10m_w^2(7s_{10} + 2s_{15}) - 4(s_{10}^2  
  - 5s_{10}s_{15} + 2s_{15}^2)) \nonumber\\&&+ m_c^2(15m_w^6 - 8s_{10}(s_{10}
  - 2s_{15})s_{15}  
  - 2m_w^4(35s_{10} + 12s_{15}) \nonumber\\&&
  + 4m_w^2(14s_{10}^2 
  + 24s_{10}s_{15} + 3s_{15}^2))) 
  + 2m_b^3m_c(53m_c^4m_w^2 \nonumber\\&&
  - 5m_w^6 - 24s_{10}^2(s_{10} + s_{15}) 
  - 2m_w^4(13s_{10} + 4s_{15}) \nonumber\\&&
  + 12m_w^2(11s_{10}^2 + s_{15}^2)
   + m_c^2(-42m_w^4 - 28m_w^2(s_{10} - 2s_{15})
   \nonumber\\&&+ 4(7s_{10}^2 + 6s_{10}s_{15} + 4s_{15}^2)))
   + m_b^2(54m_c^6m_w^2 
   + (m_w^2 \nonumber\\&&- 2s_{10})(m_w^6 + 4m_w^4(s_{10} - s_{15}) + 8s_{10}s_{15}(s_{10} + s_{15}) \nonumber\\&&
   + 4m_w^2(3s_{10}^2 - 3s_{10}s_{15} + s_{15}^2)) -4m_c^2(7m_w^6 \nonumber\\&&
   - 4m_w^4(3s_{10} + s_{15}) 
   + m_w^2(-35s_{10}^2 + 28s_{10}s_{15} - 2s_{15}^2) \nonumber\\&&
   + 4s_{10}(s_{10}^2 + s_{10}s_{15} + s_{15}^2)) +m_c^4(-23m_w^4 \nonumber\\&&
   + m_w^2(-34s_{10} + 52s_{15})
    + 4(13s_{10}^2 - 7s_{10}s_{15} + 7s_{15}^2))) \nonumber\\&&- m_c^2(13m_c^6m_w^2 + m_c^4(-23m_w^4 + m_w^2(74s_{10} + 28s_{15}) \nonumber\\&&
    + 4(3s_{10}^2 + 3s_{10}s_{15} - s_{15}^2)) -m_w^2(m_w^6 - 2m_w^4(s_{10} + 2s_{15}) \nonumber\\&&
- 16s_{10}(2s_{10}^2 + s_{15}^2)+ 4m_w^2(4s_{10}^2 + 3s_{10}s_{15} + s_{15}^2)) \nonumber\\&&
+m_c^2(11m_w^6 - 4m_w^4(17s_{10} + 6s_{15}) \nonumber\\&&+ 4m_w^2(21s_{10}^2 + 22s_{10}s_{15} + 4s_{15}^2)
+ 8(s_{10}^3 + 2s_{10}s_{15}^2))))),\nonumber
  \end{eqnarray}
  
  \begin{eqnarray}&&
   \Sigma|A_2|^2\nonumber\\
   &=&\frac{1}{m_c^2m_w^2(m_b^2 - m_c^2 + m_w^2 - 2{s_{10}} - 2{s_{15}})^4}  
   \nonumber\\&&\times(16(m_b + m_c)^2(2m_b^6m_w^2 + 8m_b^5m_cm_w^2 + m_c^4(m_w^4 - 4s_{10}^2  \nonumber\\&&
   + 4m_w^2(s_{10} - 2s_{15}) + 4s_{15}^2) + (m_w^2 - 2(s_{10} + s_{15}))^2(m_w^4\nonumber\\&&
    - 2m_w^2(s_{10} - 2s_{15})
    - 4s_{15}(s_{10} + s_{15})) -
   2m_c^2(m_w^6 \nonumber\\&&+ 4s_{10}(s_{10} + s_{15})^2 - m_w^4(s_{10} + 4s_{15}) - 4m_w^2(s_{10}^2 - s_{15}^2)) \nonumber\\&&
   + m_b^4(12m_c^2m_w^2 - m_w^4 - 4m_w^2(4s_{10} + s_{15}) + 4(3s_{10}^2 + 4s_{10}s_{15} \nonumber\\&&+ s_{15}^2)) 
   +2m_b^3m_c(12m_c^2m_w^2 - 9m_w^4 - 6m_w^2(5s_{10} + 3s_{15}) \nonumber\\&&+ 8(4s_{10}^2 + 7s_{10}s_{15} + 3s_{15}^2)) +
   2m_b^2(17m_c^4m_w^2 - m_w^6 \nonumber\\&&- 12s_{10}(s_{10} + s_{15})^2 + m_w^4(-5s_{10} + 4s_{15}) + 4m_w^2(5s_{10}^2 \nonumber\\&&
   + 4s_{10}s_{15} - s_{15}^2) + m_c^2(-16m_w^4 - 34m_w^2(s_{10} + s_{15})\nonumber\\&&
    + 44(s_{10} + s_{15})^2)) +2m_bm_c(8m_c^4m_w^2 - m_w^6 \nonumber\\&&
    - 16s_{10}(s_{10} + s_{15})^2 + m_w^4(-8s_{10} + 4s_{15}) \nonumber\\&&
    + 4m_w^2(7s_{10}^2 + 6s_{10}s_{15} - s_{15}^2) +
     m_c^2(-7m_w^4 \nonumber\\&&
     - 2m_w^2(5s_{10} + 11s_{15}) 
     + 8(2s_{10}^2 + 5s_{10}s_{15} + 3s_{15}^2))))),\nonumber
  \end{eqnarray}
  
  \begin{eqnarray}&&
   \Sigma {\rm Re}(A_1{A_2}^*)\nonumber\\
   &=&\frac{1}{m_b^3 + m_b^2m_c - m_c^3 + m_cm_w^2 + m_b(-m_c^2 + m_w^2 - 2{s_{10}})}\nonumber\\
 &&\times\frac{1}{m_c^2m_w^2(m_b^2 - m_c^2 + m_w^2 - 2{s_{10}} - 2{s_{15}})^3}
   \nonumber\\&&\times(-16(m_b + m_c)^2(4m_b^7m_w^2 + 21m_b^6m_cm_w^2 + m_b^5(38m_c^2m_w^2 \nonumber\\&&
   - m_w^4 - 2m_w^2(10s_{10} + s_{15}) + 4(3s_{10}^2 + 5s_{10}s_{15} + 2s_{15}^2)) \nonumber\\&&
   +m_b^4m_c(39m_c^2m_w^2 - 17m_w^4 - 2m_w^2(45s_{10} + 8s_{15}) \nonumber\\&&
   + 8(7s_{10}^2 + 12s_{10}s_{15} + 5s_{15}^2)) +
   m_b^2m_c(35m_c^4m_w^2 - 15m_w^6 \nonumber\\&&+ 18m_w^4(s_{10} + s_{15}) 
   - 8(5s_{10} - 2s_{15})(s_{10} + s_{15})^2 \nonumber\\&&
   + 4m_w^2(17s_{10}^2 - s_{10}s_{15} - 8s_{15}^2) 
   +m_c^2(-26m_w^4 + 52s_{10}^2 \nonumber\\&&
   + 64s_{10}s_{15} + 32s_{15}^2 - 6m_w^2(15s_{10} + 8s_{15}))) + 2m_b^3(20m_c^4m_w^2 \nonumber\\&&
   - 4m_w^6 + 3m_w^4(s_{10} + 2s_{15})
    + 2m_w^2(8s_{10}^2 - 3s_{15}^2) \nonumber\\&&
    +4(-3s_{10}^3 - 4s_{10}^2s_{15} + s_{15}^3) 
     + m_c^2(-19m_w^4 + 38s_{10}^2 \nonumber\\&&
     + 68s_{10}s_{15} + 32s_{15}^2 - m_w^2(69s_{10} + 22s_{15}))) +
   m_c(m_c^6m_w^2\nonumber\\&&
   + m_c^4(3m_w^4 - 4m_w^2s_{10} + 4s_{10}^2- 8s_{15}^2) - m_w^2(m_w^2 - 2s_{15})(m_w^4  \nonumber\\&&
   - 4m_w^2s_{10} + 4s_{10}^2 - 4s_{15}^2) +
     m_c^2(-3m_w^6 + 8s_{10}s_{15}(s_{10} + s_{15}) \nonumber\\&&
     + 2m_w^4(6s_{10} + 5s_{15}) - 4m_w^2(s_{10}^2+ 6s_{10}s_{15} + 6s_{15}^2))) \nonumber\\&& 
     +m_b(14m_c^6m_w^2 - (m_w^2 - 2s_{10})(m_w^2 - 2s_{15})(3m_w^4 \nonumber\\&&
     - 8m_w^2(s_{10} + s_{15}) + 4(s_{10} + s_{15})^2) 
     - m_c^4(m_w^4 \nonumber\\&&+ 2m_w^2(13s_{10} + 9s_{15}) 
     - 4(6s_{10}^2 + s_{10}s_{15} - 2s_{15}^2)) -
     2m_c^2(5m_w^6 \nonumber\\&&- 4m_w^4(3s_{10} + 2s_{15})
      + m_w^2(-14s_{10}^2 + 14s_{10}s_{15} + 22s_{15}^2) \nonumber\\&&
      + 4(s_{10}^3 + 2s_{10}^2s_{15} - s_{15}^3))))).\nonumber
  \end{eqnarray}


\begin{thebibliography}{9}

\bibitem{Phy} Randy M. Thurman-Keup, Ashutosh V. Kotwal, Monica Tecchio
and Aesook Byon-Wagner, Rev. Mod. Phys {\bf 73}, 267-306(2001).
\bibitem{Tev} Petroff  P, CDF DO Collaborations,  XXV Physics in Collision,
Proceedings {\bf 815}, 49-62(2006); Krasny MW, Fayette F, Placzek W,
et al, Eur.Phys.J.C {\bf 51}, 607-617(2007); Bolognesi S,  IFAE
2006: Italian Meeting on High Energy Physics 163-166(2007);
Ananthanarayan B, Patra M,Poulose P, JHEP {\bf 11}, 058(2009);
Cavaliere V, CDF and DO Collaborations, Nuovo Cimento Della Societa
Italiana Di Fisica B-general Physics Relativity Astronomy and
Mathematical Physics and Methods {\bf 123}, 757-759(2008); Rovelli
C, ATLAS and CMS Collaborations, Nuovo Cimento Della Societa
Italiana Di Fisica B-general Physics Relativity Astronomy and
Mathematical Physics and Methods {\bf 123}, 775-777(2008).
\bibitem{Pdg} Particle Data Group, K. Nakamura, et al,  J. Phys. G {\bf 37}, 075021(2010).

\bibitem{Aaltonen:2011nn}
  T.~Aaltonen {\it et al.} [ CDF Collaboration ],
  [arXiv:1104.1585 [hep-ex]].
  
\bibitem{Coo} A. M. Cooper-Sarkar, Proceedings of the Workshop on Physics at TEV Colliders,
 Les Houches 2005; hep-ph/0512228.
\bibitem{Jon} Jonathan R. Gaunt, Chun-Hay Kom, Anna Kulesza and W. James Stirling,
Eur. Phys. J. C {\bf 69}, 53-65(2010).
\bibitem{Bod} G. T. Bodwin, Eric Braaten and G. P. Lepage, Phys. Rev. D {\bf 51}, 1125-1171(1995);Erratum Phys. Rev. D {\bf55}, 5853(1997).

\bibitem{Fra} E. Braaten and T. C. Yuan, Phys. Rev. Lett {\bf 71} 1673(1993);
 Phys. Rev. D {\bf50} 3176(1994);
 C.-H. Chang and Y.-Q. Chen, Phys.Lett.B {\bf 284}, 127(1992);
 Phys. Rev. D {\bf 46}, 3845 (1992);
 Y.-Q. Chen, Phys. Rev. D {\bf 48}  5181(1993);
 T. C. Yuan, Phys. Rev. D {\bf 50} 5664(1994).
\bibitem{Bar} V. Barger, Kingman Cheung and W.-Y. Keung, Phys. Rev. D {\bf 41}, 1541-1546(1990).
\bibitem{Bra} Eric Braaten, Kingman Cheung and T. C. Yuan, Phys. Rev. D {\bf 48}, 4230-4235(1993).
\bibitem{Braa} Eric Braaten, Kingman Cheung and T. C. Yuan, Phys. Rev. D {\bf 48}, 5049-5054(1993).
\bibitem{Qiao} Cong-Feng Qiao, Chong Sheng Li and Kuang-Ta Chao, Phys. Rev. D {\bf 54}, 5606-5610(1996).
\bibitem{Cha} Chao-Hsi Chang and Xing-Gang Wu, Eur. Phys. J. C {\bf 38}, 267-276(2004).
\bibitem{Wu} Xing-Gang Wu, Phys. Lett. B {\bf 671}, 318-322(2009).
\bibitem{Eichten} E. J. Eichten and C. Quigg, Phys. Rev. D {\bf 49}, 5845-5856(1994).
\bibitem{Gao} Private communication with Y.N. Gao.
\end{thebibliography}
\end{document}